\documentclass[11pt]{article}
\usepackage[a4paper]{geometry}
\usepackage[utf8]{inputenc}
\usepackage[T1]{fontenc}
\usepackage{setspace,authblk}
\onehalfspace
\usepackage{amsmath,mathrsfs,amsfonts,amsthm}

\usepackage{cleveref}

\usepackage[english]{babel}

\usepackage{apacite}

\usepackage{graphicx}

\title{\singlespace {Fuzzy Profit Shifting: A Model for Optimal Tax-induced Transfer Pricing with Fuzzy Arm's Length Parameter}}

\author[]{Alex A.T. Rathke\thanks{School of Economics, Business and Accounting at Ribeir\~ao Preto, University of S\~ao Paulo, Brazil. E-mail: \texttt{alex.rathke@usp.br}}}

\affil[]{\textit{FEA-RP, University of S\~ao Paulo, Brazil}}
\date{\today}

\theoremstyle{plain}

\newtheorem*{theorem*}{Theorem}
\newtheorem{proposition}{Proposition}
\newtheorem*{proposition*}{Proposition}

\newtheorem*{remark*}{Remark}
\newtheorem{corollary}{Corollary}
\newtheorem*{corollary*}{Corollary}

\begin{document}

\maketitle

\begin{abstract} 

This paper proposes a model of optimal tax-induced transfer pricing with a fuzzy arm's length parameter. Fuzzy numbers provide a suitable structure for modelling the ambiguity that is intrinsic to the arm's length parameter. For the usual conditions regarding the anti-shifting mechanisms, the optimal transfer price becomes a maximising $\alpha$-cut of the fuzzy arm's length parameter. Nonetheless, we show that it is profitable for firms to choose any maximising transfer price if the probability of tax audit is sufficiently low, even if the chosen price is considered a completely non-arm's length price by tax authorities. In this case, we derive the necessary and sufficient conditions to prevent this extreme shifting strategy.

\end{abstract}

\noindent\textbf{Keywords:} fuzzy profit shifting, transfer pricing, tax evasion, tax enforcement, tax penalty.
\\
\noindent\textbf{JEL Classification:} F23, H26, K34

\section{Introduction} \label{Introduction}

Tax literature frequently draws attention to the ambiguity between a tolerant tax avoidance behaviour vs. tax evasion. This ambiguity is especially relevant on the analysis of profit shifting strategies, where multinational enterprises -- MNE carry intra-firm transactions between related parties from different jurisdictions, so to adjust the transfer prices in order to reallocate taxable profits from high-tax to low-tax locations\footnote{Existing studies provide relevant evidences of profit shifting by means of direct transfer pricing adjustments \cite{davies2018,cristea2016,bernard2006,overesch2006,bartelsman2003,clausing2003,swenson2001}.}. Anti-shifting rules require that the transfer prices comply with the so called \emph{arm's length principle} \cite{oecd2017}, which states that intra-firm prices must be consistent with ones that would have been established with independent unrelated parties. If the arm's length condition is not satisfied, tax authorities require the payment of taxes over the shifted profits, and a tax penalty usually applies.

The arm's length condition is a fuzzy concept, since independent prices are influenced by legitimate differences in transactions' conditions \cite{becker2017,eden2001,oecd2017}. It means that transfer prices are not attained to a unique true arm's length price, but rather to a range of observable parameter prices with different degrees of appropriateness with respect to the arm's length condition. In the case of a tax audit, the tax authority has to assess if the transfer prices applied by the MNE satisfy the arm's length condition, or if the deviations from the core of the arm's length range represent evidences of profit shifting. This is no more than an ambiguous decision to be taken by the tax authority, thus it implies in additional uncertainties for the MNE.

This paper derives a model for optimal tax-induced transfer pricing subjected to a fuzzy arm's length parameter. We apply \emph{fuzzy numbers}, which were first proposed by \cite{zadeh1965} and developed further by several researchers \cite{zimmermann1991,klir1995,verdegay1982}, thus to model the impact of the uncertainty that is intrinsic to the arm's length parameter over the profit-maximisation strategy. Our model follows the concealment costs approach that is traditional in profit shifting literature \cite{sandmo1972,kant1988,hines1994}, however we design it in a generalised tax condition, which allows for the maximisation analysis without constraints on the shifting direction. The model takes the arm's length parameter as a fuzzy number, therefore the maximisation object is also a fuzzy object.

Baseline analysis shows that the solution of the fuzzy maximisation object under usual conditions is a $\alpha$-cut of the fuzzy arm's length parameter, and any adjustments on the transfer price up to the optimal level provide a profit-shifting gain for the MNE. Nonetheless, we show that the MNE may completely disregard the arm's length parameter if the probability of tax audits is sufficiently low. It means that it is profitable to choose any maximising transfer price if the MNE has low chances of being audited, even if the maximising transfer price is considered a completely non-arm's length price. In this sense, we derive the necessary and sufficient conditions to prevent this extreme shifting case.

The remaining of this paper is structured as follows. Section \ref{Basics on Fuzzy Sets} presents the basic notions of fuzzy sets and fuzzy numbers. Section \ref{The Model} derives the general model. Section \ref{Optimal Transfer Pricing} solves the fuzzy maximisation object, presents the sensitivity analyses, and derives the impact of a general tax enforcement effect regarding the country-level anti-shifting variables. Section \ref{conclusion} draws some concluding comments.

\section{Basics on Fuzzy Sets} \label{Basics on Fuzzy Sets}

Fuzzy sets were first introduced by seminal paper of \cite{zadeh1965} and generalise the classical notion of crisp sets. Fuzzy sets are a collection of elements in a universe where the boundary of the set is not clearly defined. The ambiguity associated with the bounds of the fuzzy set $\tilde{A}$ in a universe $X$ is represented by a \emph{membership function} defined as $\mu_{\tilde{A}} (x) : \mathbb{R} \rightarrow [0,1]$, $x \in X$, for $\mu_{\tilde{A}} (x)$ measures the grade of membership of element $x$ in $\tilde{A}$. If the grade of membership is 0, then the element $x$ does not belong to $\tilde{A}$. If the grade of membership is 1, then the element $x$ completely belongs to $\tilde{A}$. If the grade of membership is within the interval [0,1], then the element $x$ only partially belongs to $\tilde{A}$. The fuzzy set $\tilde{A}$ is therefore characterised by the pair $\{ (x,\mu_{\tilde{A}} (x)):x \in X \}$. Two fuzzy sets $\tilde{A}$ and $\tilde{B}$ are considered equal iff $\mu_{\tilde{A}} (x) = \mu_{\tilde{B}} (x)$.

Let $\tilde{A}=\{ (x,\mu_{\tilde{A}} (x)):x \in X \}$ be a fuzzy set and define a continuous interval $\alpha \in [0,1] $. The ordinary crisp set associated with any $\alpha \in [0,1]$ is called \emph{$\alpha$-cut} of the fuzzy set $\tilde{A}$ and is defined as $A_{\alpha} = \{ x \in X : \mu_{\tilde{A}} (x) \geq \alpha \}$. We can use $\alpha$-cuts to represent intervals on fuzzy sets as

\begin{align*}
\tilde{A}_{\alpha} & = [A_{\alpha}^{\wedge},A_{\alpha}^{\vee}] \\
& = \left[ \underset{x}{\min} \{ \tilde{A} \} , \underset{x}{\max} \{ \tilde{A} \} \right] : \tilde{A}= \{ (X,\mu_{\tilde{A}} (x)) , \mu_{\tilde{A}} (x) \geq \alpha \} .
\end{align*}

The sets $A_{\alpha}$, $\alpha \in [0,1]$ refer to a decreasing succession of subsets continua; $\alpha_{1} \geq \alpha_{2} \Leftrightarrow A_{\alpha_{1}} \subseteq A_{\alpha_{2}}$, $\alpha_{1}$, $\alpha_{2} \in [0,1]$ \cite{klir1995}.

\begin{theorem*}
(Representation Theorem - \cite{klir1995,zimmermann1991,verdegay1982}\footnote{\cite{klir1995} analyse this theorem in a set of three \emph{Decomposition Theorems} for representation of fuzzy sets by means of their $\alpha$-cuts.}) For a fuzzy set $\tilde{A}$ and its $\alpha$-cuts $A_{\alpha}$, $\alpha \in [0,1]$, we have

\begin{equation*}
\tilde{A} = \underset{\alpha \in [0,1]}{\bigcup} \alpha \cdot A_{\alpha} .
\end{equation*} 

If the membership function $\mu_{A_{\alpha}} (x)$ is defined as the characteristic function of the set $A_{\alpha}$

\begin{equation*}
\mu_{A_{\alpha}} (x) = \left\{
 \begin{array}{ll}
 1, & \text{iff } x \in A_{\alpha}\\
 0, & \text{otherwise}\\
 \end{array} \right.
\end{equation*}

\noindent the membership function of the fuzzy set $\tilde{A}$ can be expressed as the characteristic function of its $\alpha$-cuts as

\begin{equation*}
\mu_{\tilde{A}} (x) = \underset{\alpha \in [0,1]}{\sup} \min{(\alpha , \mu_{A_{\alpha}} (x))} .
\end{equation*} 
\end{theorem*} \hfill $\triangle$

A fuzzy set $\tilde{A}$ is \emph{convex} iff its $\alpha$-cuts are convex. Equivalently, $\tilde{A}$ is convex iff $\forall x_{1}$, $x_{2} \in X$, $\lambda \in [0,1] : \mu_{\tilde{A}} (\lambda x_{1} + (1 - \lambda) x_{2}) \geq \min{(\mu_{\tilde{A}} (x_{1}),\mu_{\tilde{A}} (x_{2}))}$. A fuzzy set $\tilde{A}$ is \emph{normalised} iff $\sup_{x \in X} \mu_{\tilde{A}} = 1$.

A \emph{fuzzy number} is a special case of a fuzzy set on the real line that is both convex and normalized. Its membership function is piecewise continuous and $\exists x_{0} \in \mathbb{R} : \mu_{\tilde{A}} (x_{0}) = 1$ is called its \emph{mode}. Since fuzzy sets are completely defined by their corresponding membership functions, we refer to a fuzzy number as the set $\tilde{A}$ as well as the membership function $\mu_{\tilde{A}} (x)$ hereinafter. For a sequence of real numbers $x^{\wedge} \leq \bar{x}^{\wedge} \leq \bar{x}^{\vee} \leq x^{\vee} \in \mathbb{R}$, the fuzzy number $\tilde{A}$ satisfies the following:
\begin{enumerate}
 \item[a.] $\mu_{\tilde{A}} (x) = 0$ for each $x \notin [x^{\wedge},x^{\vee}]$ ;
 \item[b.] $\mu_{\tilde{A}} (x)$ is non-decreasing in $[x^{\wedge},\bar{x}^{\wedge}]$ and non-increasing in $[\bar{x}^{\vee},x^{\vee}]$ ;
 \item[c.] $\mu_{\tilde{A}} (x) = 1$ for each $x \in [\bar{x}^{\wedge},\bar{x}^{\vee}]$ ;
\end{enumerate}

\noindent where $[\bar{x}^{\wedge},\bar{x}^{\vee}]$ is the mode of the fuzzy number, $[x^{\wedge},\bar{x}^{\wedge}]$ is the interval on the lower side of the mode with width $\bar{x}^{\wedge} - x^{\wedge}$, and $[\bar{x}^{\vee},x^{\vee}]$ is the interval on the upper side of the mode with width $x^{\vee} - \bar{x}^{\vee}$. A fuzzy number $\tilde{A}$ is of the $\mathcal{LR}$\emph{-type} if it can be parametrised by shape functions $f^{\wedge}(\cdot)$ and $f^{\vee}(\cdot)$  on the lower and upper sides of the mode respectively\footnote{Literature commonly refer to the \emph{left} and \emph{right} sides of the mode $\mu_{\tilde{A}} (x) = 1$, i.e. though the origin of the term $\mathcal{LR}$-type with shape functions $\mathcal{L}(\cdot)$ and $\mathcal{R}(\cdot)$.}. A \emph{plane} fuzzy number satisfies $\exists (\bar{x}^{\wedge},\bar{x}^{\vee}) \in \mathbb{R}$, $\bar{x}^{\wedge} < \bar{x}^{\vee} : \forall x \in [\bar{x}^{\wedge},\bar{x}^{\vee}] \rightarrow \mu_{\tilde{A}} (x) = 1$, i.e. its mode is a non-empty interval with more than one element \cite{klir1995,zimmermann1991}. A fuzzy number is called a \emph{trapezoidal} fuzzy number iff it takes the form

\begin{equation*}
\mu_{\tilde{A}} (x) = \left\{
 \begin{array}{ll}
 \dfrac{x-x^{\wedge}}{\bar{x}^{\wedge}-x^{\wedge}}, & x^{\wedge} \leq x \leq \bar{x}^{\wedge} \\
 \\
 1, & \bar{x}^{\wedge} \leq x \leq \bar{x}^{\vee} \\
 \\
 \dfrac{x^{\vee}-x}{x^{\vee}-\bar{x}^{\vee}}, & \bar{x}^{\vee} \leq x \leq x^{\vee} \\
 \\
 0, & \text{otherwise.}\\
 \end{array} \right.
\end{equation*}

A fuzzy number is called a \emph{triangular} fuzzy number iff it takes the form

\begin{equation*}
\mu_{\tilde{B}} (x) = \left\{
 \begin{array}{ll}
 \dfrac{x-x^{\wedge}}{\bar{x}^{\wedge}-x^{\wedge}}, & x^{\wedge} \leq x \leq \bar{x}^{\wedge} \\
 \\
 \dfrac{x^{\vee}-x}{x^{\vee}-\bar{x}^{\wedge}}, & \bar{x}^{\wedge} \leq x \leq x^{\vee} \\
 \\
 0, & \text{otherwise.}\\
 \end{array} \right.
\end{equation*}

\begin{figure}[h]
  \includegraphics[width=\textwidth]{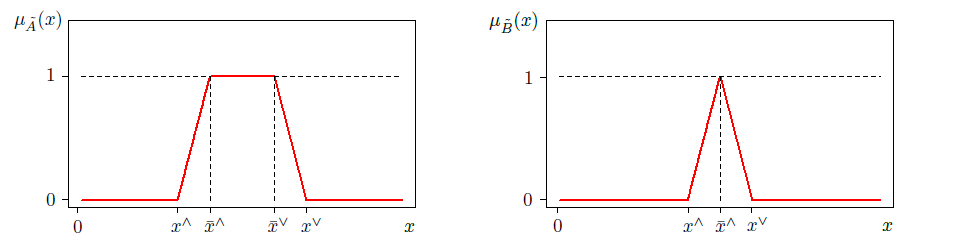}
  \caption{Examples of fuzzy numbers: A symmetric trapezoidal fuzzy number $\tilde{A}$ on the left and a symmetric triangular fuzzy number $\tilde{B}$ on the right. Both $\tilde{A}$ and $\tilde{B}$ are special fuzzy numbers of the $\mathcal{LR}$-type.}
  \label{fig:fuzzynumbers}
\end{figure}

\bigskip

Figure \ref{fig:fuzzynumbers} shows examples of trapezoidal and triangular fuzzy numbers. It is clear that a trapezoidal fuzzy number is an instance of plane fuzzy number, and a triangular fuzzy number refers to a trapezoidal fuzzy number with $\bar{x}^{\wedge}=\bar{x}^{\vee}$.

\section{The Model} \label{The Model}

In this section, we derive a model to analyse the optimal tax-induced transfer pricing. We first set the baseline net profit function for the MNE, then we derive the specification of the fuzzy profit shifting optimisation.

\subsection{Baseline Profit Design} \label{Baseline Profit Design}

Consider a vertically integrated MNE with two divisions, the parent company located in Country 1 and a wholly owned subsidiary located in Country 2, $i = \{ 1,2 \}$. Both divisions\footnote{For simplification, we apply subscript $i$ for the reference of both countries and to each MNE's divisions hereinafter.} produce outputs $x_{i}$ under costs $C_{i}(x_{i})$, bringing revenues $R_{i}(s_{i})$ based on domestic sales $s_{i}(x_{i})$. Parent firm also exports a portion $m$ of its output to subsidiary in Country 2, regarding a single type of product, charging a transfer price $p$ established by means of exclusive self-discretion of MNE's central management. We set $m=m(s_{2})$ and $\partial{m}/\partial{s_{2}} > 0$, thus intra-firm output $m$ depends on the market demand for final product in Country 2. The pre-tax profits of both divisions are

\begin{align*}
\pi_{1}=R_{1}(s_{1})-C_{1}(s_{1}+m)+pm; \\
\pi_{2}=R_{2}(s_{2})-C_{2}(s_{2}-m)-pm.
\end{align*}

Country 1 applies the source principle on taxation of foreign profits, and we assume no incremental operational cost on transferring internal output $m$ to division 2, i.e. $\partial{C_{i}(m,x_{i})}/\partial{m}=\partial{C_{i}(m,x_{i})}/\partial{x_{i}}$. For an income tax rate $\tau_{i} \in [0,1]$ in each country, MNE's global net profits are $\Pi(\tau_{i},s_{i},p,m)=(1-\tau_{1})\pi_{1}+(1-\tau_{2})\pi_{2}$. Profit shifting incentives arise when tax rates between divisions are different, $\tau_{1} \neq \tau_{2}$, and total net profit $\Pi(\cdot)$ increases when MNE is able to choose a specific transfer price $p$ so profits are transferred from the high-tax country to the low-tax country. The condition $\partial{\Pi(\cdot)}/\partial{p}=(\tau_{2}-\tau_{1})m$ implies the following two cases:


\begin{equation} \label{tp cases}
\begin{array}{ll}
\text{Low Transfer Price case - LTP:} & \quad \tau_{2} < \tau_{1} \rightarrow \dfrac{\partial{\Pi(\cdot)}}{\partial{p}} < 0; \\
\\
\text{High Transfer Price case - HTP:} & \quad \tau_{2} > \tau_{1} \rightarrow \dfrac{\partial{\Pi(\cdot)}}{\partial{p}} > 0 .
\end{array} 
\end{equation}

In the LTP case, the MNE has incentives to shift profits from division 1 to division 2 by choosing a low transfer price $p$, thus harming tax revenues in Country 1. In the HTP case, MNE chooses a high price $p$ so to shift profits to the opposite direction, thus harming Country 2.


\subsection{Fuzzifying the Arm's Lenght Price} \label{Fuzzifying the Arm's Lenght Price}

Assume that both countries impose a non-negligible and non-deductible tax penalty $z_{i} >0$ if profit shifting is detected, which is computed as a portion of the amount of evaded taxes. It means that the tax penalty $z_{i}$ is imposed if the harmed Country $i$ observes that the transfer price $p$ is different from a parameter price $\bar{p}$ established under arm's length conditions\footnote{The transfer pricing guidelines prepared by \cite{oecd2017} have become the main criterion adopted by most countries worldwide for evaluation of intra-firm prices. The guidelines are built on the basis of the arm's length principle as the fundamentals for tax-compliant transfer pricing.} and this price gap results in the outflow of taxable profits from Country $i$. The parameter of an arm's length price is a fuzzy concept, since independent prices vary according to legitimate differences in transactions' conditions. Therefore, countries rather observe a fuzzy set of parameter prices $\tilde{P}$, all of which have different degrees of appropriateness with respect to the arm's length principle\footnote{In this line, anti-shifting rules usually establish an arm's length range of appropriate transfer prices. The arm's length range is usually set as an interquartile range within the complete set of comparable prices \cite{oecd2017}.}.

Define the fuzzy set of arm's length prices $\tilde{P} = \{ (p_{j},\mu_{\tilde{P}} (p_{j})): p_{j} \in P \}$, $j \neq i$, $P \in \mathbb{R}_{+}$, where $P$ is the universe of all observable independent prices, universe $P$ is convex, and $\mu_{\tilde{P}} (p_{j})$ is the membership function of the fuzzy set $\tilde{P}$. For a sequence of independent prices $p^{\wedge} \leq \bar{p}^{\wedge} \leq \bar{p}^{\vee} \leq p^{\vee} \in P$, the fuzzy set $\tilde{P}$ satisfies the usual conditions

\begin{equation} \label{exclusion condition}
\mu_{\tilde{P}} (p_{j}) = 0 \text{ for } \forall p_{j} \notin [p^{\wedge},p^{\vee}] ;
\end{equation}
\begin{equation} \label{non-decreasing condition}
\mu_{\tilde{P}} (p_{j}) \text{ is non-decreasing for } \forall p_{j} \in [p^{\wedge},\bar{p}^{\wedge}] ;
\end{equation}
\begin{equation} \label{non-increasing condition}
\mu_{\tilde{P}} (p_{j}) \text{ is non-increasing for } \forall p_{j} \in [\bar{p}^{\vee},p^{\vee}] .
\end{equation}

The mode of the fuzzy set $\tilde{P}$ satisfies $\forall p_{j} \in P : \mu_{\tilde{P}} (p_{j}) = 1$, which provides the interval of prices that completely satisfy the arm's length principle, $\mu_{\tilde{P}} (p_{j}) = 1$ for $\forall p_{j} \in [\bar{p}^{\wedge},\bar{p}^{\vee}]$. Hence, the choice of any strict parameter price $\bar{p}$ must lie within the interval of prices that define the mode of the fuzzy set $\tilde{P}$, i.e. $\bar{p} \in [\bar{p}^{\wedge},\bar{p}^{\vee}]$. Eq. \ref{exclusion condition} defines the limiting interval $[p^{\wedge},p^{\vee}]$ out of which any price $p$ is considered a completely non-arm's length price.

Under these conditions, the fuzzy set $\tilde{P}$ becomes a fuzzy number of the $\mathcal{LR}$-type. Call $\tilde{P}$ the fuzzy arm's length price. We define a standard membership function of the fuzzy number $\tilde{P}$ as follows:

\begin{equation} \label{fuzzy arms length price}
\mu_{\tilde{P}} (p_{j}) = \left\{
 \begin{array}{ll}
 f^{\wedge} \left( \dfrac{p_{j} - p^{\wedge}}{\bar{p}^{\wedge} - p^{\wedge}} \right), & p^{\wedge} \leq p_{j} \leq \bar{p}^{\wedge} \\
 \\
 1, & \bar{p}^{\wedge} \leq p_{j} \leq \bar{p}^{\vee} \\
 \\
 f^{\vee} \left( \dfrac{p_{j} - p^{\vee}}{\bar{p}^{\vee} - p^{\vee}} \right), & \bar{p}^{\vee} \leq p_{j} \leq p^{\vee} \\
 \\
 0, & \text{otherwise}\\
 \end{array} \right.
\end{equation}

\noindent with both functions $f^{\wedge}(\cdot)$ and $f^{\wedge}(\cdot)$ monotone continuous. In Eq. \ref{fuzzy arms length price}, we allow for the fuzzy arm's length price $\tilde{P}$ to be asymmetric. This asymmetry may be due to a difference in the widths $\bar{p}^{\wedge} - p^{\wedge}$ and $p^{\vee} - \bar{p}^{\vee}$ on the lower and upper sides of the fuzzy number $\tilde{P}$ respectively, as well as for differences in grades of membership denoted by functions $f^{\wedge}(\cdot)$ and $f^{\vee}(\cdot)$. In effect, the asymmetry in the fuzzy arm's length price $\tilde{P}$ is useful to describe how Countries 1 and 2 differ in their tolerance for a transfer price $p$ farther from the parameter price $\bar{p}$.

For the LTP case in Eq. \ref{tp cases}, Country 1 is less tolerant with respect to a low transfer price close to $p^{\wedge}$, while it accepts prices near or higher than the parameter price $\bar{p}$. Therefore, Country 1 is only concerned with the lower side $f^{\wedge}(\cdot)$ of the fuzzy arm's length price $\tilde{P}$. The opposite occurs for the HTP case in Eq. \ref{tp cases}, since Country 2 is only concerned with the higher side $f^{\vee}(\cdot)$ of $\tilde{P}$. If we divide the fuzzy arm's length price $\tilde{P}$ into two membership sections with respect to lower side $f^{\wedge}(\cdot)$ and upper side $f^{\vee}(\cdot)$, we obtain two fuzzy numbers $\tilde{P}^{\wedge}$ and $\tilde{P}^{\vee}$ satisfying the additional conditions:

\begin{equation} \label{lower fuzzy arms length price}
\mu_{\tilde{P}^{\wedge}} (p_{j}) = \left\{
 \begin{array}{ll}
 \mu_{\tilde{P}} (p_{j}), & p_{j} \leq \bar{p}^{\vee} \\
 1, & p_{j} > \bar{p}^{\vee} .
 \end{array} \right.
\end{equation}

\begin{equation} \label{upper fuzzy arms length price}
\mu_{\tilde{P}^{\vee}} (p_{j}) = \left\{
 \begin{array}{ll}
 \mu_{\tilde{P}} (p_{j}), & p_{j} \geq \bar{p}^{\wedge} \\
 1, & p_{j} < \bar{p}^{\wedge} .
 \end{array} \right.
\end{equation}

\begin{equation} \label{intersection fuzzy arms length price}
\tilde{P} = \tilde{P}^{\wedge} \cap \tilde{P}^{\vee} .
\end{equation}

It is clear that the fuzzy numbers $\tilde{P}^{\wedge}$ and $\tilde{P}^{\vee}$ refer to the fuzzy arm's length prices taken into account by Countries 1 and 2 respectively\footnote{It is also clear that the fuzzy numbers $\tilde{P}^{\wedge}$ and $\tilde{P}^{\vee}$ are of the $\mathcal{L}$-type and $\mathcal{R}$-type respectively.}. We indicate the standard form of the fuzzy arm's length prices satisfying conditions in Eq. \ref{lower fuzzy arms length price}-\ref{intersection fuzzy arms length price} as $\tilde{P}^{c}$, $c=\{ \wedge,\vee \} $. The mode of the fuzzy numbers $\tilde{P}^{c}$ satisfies the standard condition $\forall p_{j} \in P : \mu_{\tilde{P}^{c}}(p_{j}) = 1$. The bound of the mode of the fuzzy numbers $\tilde{P}^{c}$ is defined in standard form\footnote{The bound $\bar{p}^{c}$ of the mode of the fuzzy number $\tilde{P}^{c}$ can be defined as
\begin{equation*}
\bar{p}^{c} : \mu_{\tilde{P}^{c}}(\bar{p}^{c} + \Delta p) < 1 , \lim_{\Delta p \rightarrow 0} \mu_{\tilde{P}^{c}}(\bar{p}^{c} + \Delta p) = 1 
\end{equation*}
\noindent with deviation $\Delta p \in \mathbb{R}$.} as $\bar{p}^{c}$. Hence, both profit shifting cases in Eq. \ref{tp cases} imply LTP $\rightarrow \{ i=1, c=\wedge \}$, HTP $\rightarrow \{ i=2, c=\vee \}$. 

\subsection{Tax Audits and Tax Penalties} \label{Tax Audits and Tax Penalties}

Both countries perform tax audits in order to prevent the profit shifting. In the universe of all taxpayers, we assume that countries are not able to continuously observe all MNE in absolute completeness, but they have to \emph{ex ante} select which MNE are going to be audited. In special, both countries have no prior knowledge about the existence of intra-firm transactions $pm$, though this knowledge depends on an initial pick. Following \cite{levaggi2013}, we set the audit selection in Country $i$ as a Poisson process with intensity rate $\lambda_{i} > 0$ homogeneous through the total period determined in the legal statute of limitations. Rate $\lambda_{i}$ refers to the tax audit intensity in Country $i$. If the MNE is selected, Country $i$ will observe $pm$, thus triggering a chance for tax penalty $z_{i}$. 

If the number of tax audits performed by Country $i$ is $q \in \mathbb{N}$, the probability of exact $q=k$ tax audits is \smash{$\mathbb{P}(q=k,\lambda_{i}) = \lambda_{i}^{k}e^{-\lambda_{i}} / k!$}. Furthermore, the cumulative probability of Country $i$ to perform up to $k$ audits, $\mathbb{P}(0 \leq q \leq k, \lambda_{i})$ is computed as

\begin{equation} \label{prob tax audit}
\mathbb{P}(0 \leq q \leq k, \lambda_{i}) = \sum_{q=0}^{k} \mathbb{P}(q,\lambda_{i}) = \dfrac{\Gamma(k+1,\lambda_{i})}{\Gamma(k+1)}
\end{equation}

\noindent where $\Gamma(k)$ is the gamma function and $\Gamma(k,\lambda)$ is the upper gamma function\footnote{Derivation of Eq. \ref{prob tax audit} in Appendix.}. Remark that no penalisation will be imposed if there is no tax audit, $q=0$. Moreover, even with an estimate of the number of tax audits $\mathbb{E}(q=k, \lambda_{i}) = \lambda_{i}$, the MNE can be selected under any number $q$ different from $k$. In summary, MNE has a chance of being selected for tax audit if Country $i$ performs at least one audit. Therefore, the total probability of tax audit for the MNE is 

\begin{equation} \label{total prob tax audit}
\mathbb{P}(q > 0, \lambda_{i}) = 1 - \mathbb{P}(q = 0, \lambda_{i}) = 1 - \dfrac{\Gamma(1,\lambda_{i})}{\Gamma(1)} = 1 - e^{-\lambda_{i}} .
\end{equation}

In the case of audit selection, Country $i$ observes the intra-firm transactions $pm$ and compares the transfer price $p$ with the arm's length parameter $\bar{p}$. If the harmed Country $i$ concludes that the MNE is shifting taxable profits away, the MNE is required to pay the amount of evaded taxes plus a penalty $z_{i}$ levied over this amount. In this case, tax penalty is computed as $Z_{i}(z_{i},\tau_{i},p,\bar{p},m) = (1 + z_{i}) \cdot sgn(\tau_{2} - \tau_{1}) \tau_{i} \cdot (p- \bar{p})m \geq 0$, where $sgn(\cdot)$ is the sign function and tax rates are non-negative, $\tau_{i} \in [0,1]$ . Observe that the total tax penalty is non-negative $Z_{i}(\cdot) \geq 0$ for both LTP and HTP cases\footnote{Total tax penalty $Z_{i}(\cdot) \geq 0 $ is non-negative since the signs of both the tax differential $\tau_{2} - \tau_{1}$ and the price gap $p- \bar{p}$ carry information about the shifting direction; HTP implies $\tau_{2} - \tau_{1} >0$, $p- \bar{p} >0$, while LTP implies $\tau_{2} - \tau_{1} <0$, $p- \bar{p} <0$.}.

Nonetheless, the assessment of the transfer price $p$ by Country $i$ is based on the fuzzy arm's length parameter $\tilde{P}^{c}$, $c = \{\wedge,\vee \}$. Formally, this assessment is made by taking the fuzzy number $\tilde{P}^{c} = \{ (p_{j},\mu_{\tilde{P}^{c}} (p_{j})): p_{j} \in P \}$ and setting the equality $p = p_{j}$. The result is a fuzzy price gap $\Delta p = \widetilde{p - \bar{p}^{c}}$, where $\bar{p}^{c}$ is the bound of the mode of the fuzzy number $\tilde{P}^{c}$. The fuzzy price gap $\Delta p$ is defined such as to satisfy the condition $p = \{(\bar{p}^{c}+\Delta p,\mu_{\tilde{P}^{c}}(\bar{p}^{c}+\Delta p)): p \in P \} $. For the harmed Country $i$, profit shifting may exist iff $\mu_{\tilde{P}^{c}}(p) < 1$, i.e. iff the fuzzy price gap $\Delta p$ pushes the transfer price $p$ away from the mode of $\tilde{P}^{c}$, $\forall p_{j} : \mu_{\tilde{P}^{c}}(p_{j}) = 1$. In this case, the original tax penalty $Z_{i}(\cdot) \geq 0$ turns into a fuzzy tax penalty in the following standard form:

\begin{equation} \label{fuzzy tax penalty}
\tilde{Z}_{i}(z_{i},\tau_{i},\Delta p,m) = \left\{
 \begin{array}{ll}
 0, & \mu_{\tilde{P}^{c}}(p) = 1 \\
 (1 + z_{i}) \cdot sgn(\tau_{2} - \tau_{1}) \tau_{i} \cdot (\widetilde{p - \bar{p}^{c}}) m, & \text{otherwise.} \\
 \end{array} \right.
\end{equation}

It means that the harmed Country $i$ has the task to assess if the price gap $\Delta p$ is a tolerable variance under the fuzzy arm's length conditions or if it is an evidence of profit shifting.

\section{Optimal Transfer Pricing} \label{Optimal Transfer Pricing}

The MNE aims choose a transfer price $p$ so to maximise global net profits $\Pi(\cdot)$, however it faces the chance of tax penalisation if the harmed Country $i$ finds out the existence of intra-firm transactions $pm$ and decides that it represents a profit shifting strategy. In this line, assuming that the optimal transfer price $p^{*}$ implies $\mu_{\tilde{P}^{c}}(p^{*}) < 1$, the MNE has a maximisation object specified as follows:


\begin{equation} \label{max objective}
\begin{array}{rcl}
 \smash{\underset{p \in P}{\max}} \quad \mathbb{E}(\tilde{\Pi}(\cdot)) &=& \Pi(\tau_{i},s_{i},p,m) - \mathbb{E} (\tilde{Z}_{i}(z_{i},\tau_{i},\Delta p,m)) \\
\\
&=& (1-\tau_{1})\pi_{1}+(1-\tau_{2})\pi_{2} \\
&&- (1 - e^{- \lambda_{i}}) \cdot (1 + z_{i}) \cdot sgn(\tau_{2} - \tau_{1}) \tau_{i} \cdot (\widetilde{p - \bar{p}^{c}}) m . \\
 \end{array}
\end{equation}

Since the expected tax penalty $\mathbb{E}(\tilde{Z}_{i}(\cdot))$ is a fuzzy number, objective function in Eq. \ref{max objective} becomes a fuzzy objective, and profit maximisation must take into account the fuzziness of the price gap $\Delta p = \widetilde{p - \bar{p}^{c}}$.

Conditions in Eq. \ref{lower fuzzy arms length price}-\ref{intersection fuzzy arms length price} show that the standard-form fuzzy arm's length price $\tilde{P}^{c}$ represents a one-to-one and onto correspondence $\mu_{\tilde{P}^{c}}(p_{j}) : \mathbb{R} \rightarrow [0,1]$ with respect to the closed interval of interest $p_{j} \in [p^{c},\bar{p}^{c}]$. Therefore, we solve Eq. \ref{max objective} by applying the procedure for fuzzy optimisation developed in the classical work of \cite{verdegay1982}. 

For the membership function $\mu_{\tilde{P}^{c}} (p_{j})$, $p_{j} \in [p^{c},\bar{p}^{c}]$, the corresponding $\alpha$-cuts are $P_{\alpha}^{c} = \{ p_{j} \in [p^{c},\bar{p}^{c}] : \mu_{\tilde{P}^{c}} (p_{j}) \geq \alpha \}$. From the representation theorem for fuzzy sets, Eq. \ref{max objective} is expressed in the following parametric form:

\begin{equation} \label{max objective parametric}
\begin{array}{rcl}
 \smash{\underset{\substack{\alpha \in [0,1] \\ p \in P_{\alpha}^{c}}}{\max}} \quad \mathbb{E}(\tilde{\Pi}(\cdot)) &=& (1-\tau_{1})\pi_{1}+(1-\tau_{2})\pi_{2} \\
&&- (1 - e^{- \lambda_{i}}) \cdot (1 + z_{i}) \cdot sgn(\tau_{2} - \tau_{1}) \tau_{i} \cdot (p - \bar{p}^{c})f(\alpha)m \\
 \end{array}
\end{equation}

\noindent with $\alpha \in [0,1]$, where $f(\alpha) : [0,1] \rightarrow P \in \mathbb{R}_{+}$, \smash{$f(\alpha) = \mu_{\tilde{P}^{c}}^{-1} (\alpha)$} is the inverse function of the membership function $\mu_{\tilde{P}^{c}} (p_{j})$. Simply stated, if the solution of Eq. \ref{max objective parametric} is $p^{*}(\alpha)$, then the solution of Eq. \ref{max objective} is the fuzzy set $p^{*} = \{ (p(\alpha), \alpha) \}$. Hence, profit maximisation in Eq. \ref{max objective} resumes to find the optimal $\alpha$-cut defined by $P_{= \alpha}^{c} = \{ p^{*}(\alpha) \in [p^{c},\bar{p}^{c}] : \mu_{\tilde{P}^{c}} (p^{*}(\alpha)) = \alpha \}$ at the membership grade $\mu_{\tilde{P}^{c}} (p^{*}(\alpha)) = \alpha$. 

Based on the general Stone-Weierstrass approximation, assume that the standard-form shape function $f^{c}(\cdot)$ in Eq. \ref{fuzzy arms length price} can be defined as a simple power function

\begin{equation} \label{shape function}
f^{c} \left( \dfrac{p-p^{c}}{\bar{p}^{c}-p^{c}} \right) = \left( \dfrac{p-p^{c}}{\bar{p}^{c}-p^{c}} \right)^{\gamma_{i}}
\end{equation}

\noindent with $\gamma_{i} \in (0,1]$ as a regularised parameter for the tolerance of Country $i$ regarding fuzziness in the arm's length price, e.g. a slacken tax assessment by Country $i$ implies $\gamma_{i} \rightarrow 0$, while a tighten tax assessment implies $\gamma_{i} \rightarrow 1$. Eq. \ref{shape function} provides a smooth variation in membership grade as transfer price $p$ gets farther from the bound of the mode $\bar{p}^{c}$. For the interval of interest $p \in [p^{c},\bar{p}^{c}]$, parametric optimisation in Eq. \ref{max objective parametric} then becomes

\begin{equation} \label{max objective parametric in p}
\begin{array}{rcl}
 \smash{\underset{p \in [p^{c},\bar{p}^{c}]}{\max}} \quad \mathbb{E}(\tilde{\Pi}(\cdot)) &=& (1-\tau_{1})\pi_{1}+(1-\tau_{2})\pi_{2} \\
&&- (1 - e^{- \lambda_{i}}) \cdot (1 + z_{i}) \cdot sgn(\tau_{2} - \tau_{1}) \tau_{i} \cdot (p - \bar{p}^{c}) \cdot \mu_{\tilde{P}^{c}}^{-1} (\alpha)m \\
\\
&=& (1-\tau_{1})\pi_{1}+(1-\tau_{2})\pi_{2} \\
&& - (1 - e^{- \lambda_{i}}) \cdot (1 + z_{i}) \cdot sgn(\tau_{2} - \tau_{1}) \tau_{i} \cdot (p - \bar{p}^{c}) \smash{\left(1- \left( \dfrac{p-p^{c}}{\bar{p}^{c}-p^{c}} \right) \right)^{\frac{1}{\gamma_{i}}}} m \\
\\
&=& (1-\tau_{1})\pi_{1}+(1-\tau_{2})\pi_{2} \\
&& - (1 - e^{- \lambda_{i}}) \cdot (1 + z_{i}) \cdot sgn(\tau_{2} - \tau_{1}) \tau_{i} \cdot (p - \bar{p}^{c}) \smash{\left( \dfrac{p-\bar{p}^{c}}{p^{c}-\bar{p}^{c}} \right)^{\frac{1}{\gamma_{i}}}} m . \\
 \end{array}
\end{equation}

Now we have the expected net profits $\mathbb{E}(\tilde{\Pi}(\cdot))$ specified completely in terms of the transfer price\footnote{Parametric form in Eq. \ref{max objective parametric in p} is possible since the arm's length parameters $p^{c},\bar{p}^{c} \in P$ are exogenous with respect to $\Pi(\cdot)$ and $\tilde{Z}(\cdot)$.} $p$. Differentiating Eq. \ref{max objective parametric in p} with respect to $p$ and solving, we obtain the solution

\begin{equation} \label{optimal p}
 \begin{array}{rcl}
\smash{\dfrac{\partial \mathbb{E}(\tilde{\Pi}(\cdot))}{\partial p}}&=& (\tau_{2} - \tau_{1})m - (1 - e^{- \lambda_{i}}) \cdot (1+z_{i}) \cdot sgn(\tau_{2} - \tau_{1})\tau_{i} \cdot \smash{\left( 1 + \dfrac{1}{\gamma_{i}} \right)} \smash{\left( \dfrac{p-\bar{p}^{c}}{p^{c} - \bar{p}^{c}} \right)^{\frac{1}{\gamma_{i}}}} m = 0 ; \\
\\
\\
p^{*}&=& \bar{p}^{c} + \left( \dfrac{\tau_{2} - \tau_{1}}{(1 - e^{- \lambda_{i}}) \cdot (1+z_{i}) \cdot sgn(\tau_{2} - \tau_{1})\tau_{i} \cdot \left( 1 + \frac{1}{\gamma_{i}} \right)} \right)^{\gamma_{i}} (p^{c} - \bar{p}^{c}) \\
\\
&=& \bar{p}^{c} + \left( \dfrac{|\tau_{2} - \tau_{1}|}{(1 - e^{- \lambda_{i}}) \cdot (1+z_{i}) \cdot \tau_{i} \cdot \left( 1 + \frac{1}{\gamma_{i}} \right)} \right)^{\gamma_{i}} (p^{c} - \bar{p}^{c}) \\
\\
&=& P_{= \alpha}^{c} = \{ p^{*} \in [p^{c},\bar{p}^{c}] : \mu_{\tilde{P}^{c}} (p^{*}) = \alpha \}
 \end{array}
\end{equation}

\noindent with $|\cdot| : \mathbb{R} \rightarrow \mathbb{R}_{+}$ as the absolute value function\footnote{The following property is applied: for any real number $\forall x \in \mathbb{R}$, $x$ satisfies
\begin{equation*}
x = sgn(x)\cdot|x| \rightarrow |x| = \frac{x}{sgn(x)} .
\end{equation*}
}. Eq. \ref{optimal p} shows that the optimal transfer price $p^{*}$ is represented as a maximising $\alpha$-cut of the fuzzy arm's length price $\tilde{P}^{c}$ defined as $P_{= \alpha}^{c} = \{ p^{*} \in [p^{c},\bar{p}^{c}] : \mu_{\tilde{P}^{c}} (p^{*}) = \alpha \}$, i.e. the optimal price gap $\Delta p^{*} = p^{*} - \bar{p}^{c}$ is a share of the price difference $p^{c} - \bar{p}^{c}$. This $\alpha$-cut is represented by a share function over the interval $[p^{c}, \bar{p}^{c}]$, which is measured as the magnitude of the profit shifting incentive $|\tau_{2} - \tau_{1}|$ adjusted by the marginal expected penalisation effect $(1 - e^{-\lambda_{i}}) \cdot (1 + z_{i}) \cdot \tau_{i}$. The slope of this share is the same as of the shape function in Eq. \ref{shape function} by means of the exponent $\gamma_{i}$. It also has an adjustment equal to $(\gamma_{i} + 1)/\gamma_{i}$, which derives from the endogenous specification of the fuzzy arm's length price $\tilde{P}^{c}$ in terms of $p$ within the expected tax penalty in Eq. \ref{max objective parametric in p}\footnote{More specifically, the transfer price $p$ affects both the transfer price gap $\Delta p = p - \bar{p}^{c}$ and the membership relation $\mu_{\tilde{P}^{c}}(p)$ specified by the shape function in Eq. \ref{shape function}, for the combined marginal effect on $\tilde{Z}(\cdot)$ becomes $(\gamma_{i} + 1)/\gamma_{i}$. On the other hand, the transfer price $p$ affects marginally the net profits $\Pi(\cdot)$ in a direct way. The total effect on the expected net profits is equal to
\begin{equation*}
  1 - \dfrac{1}{\frac{\gamma_{1} + 1}{\gamma_{1}}} = \dfrac{1}{1 + \gamma_{i}} .
\end{equation*}
}. Moreover, the amount of intra-firm output $m$ does not affect the optimal transfer price $p^{*}$ in the model, i.e. it refers to the application of the pure comparable uncontrolled price -- CUP method\footnote{Literature indicates that profit shifting detection is more effective if tax audits focus on the amount of intra-firm transfers $pm$ rather than on transfer prices only \cite{nielsen2014}. Nonetheless, anti-shifting rules require the application of the arm's length principle solely for the establishment of transfer prices $p$, i.e. there are no current requirements for an "arm's length quantity" -- say "$\bar{m}$", and tax authorities bear no arguments against any intra-firm output $m$ as long as the transfer price is equal to the arm's length price, $p = \bar{p}$. Eq. \ref{optimal p} reflects this condition.} \cite{oecd2017}. Sufficient joint conditions for non-zero optimal $\Delta p^{*}$ are $\tau_{1} \neq \tau_{2}$, $z_{i} < \infty$ and $p^{c} \neq \bar{p}^{c}$.

Recall that the two profit shifting cases in Eq. \ref{tp cases} imply LTP $\rightarrow \{ i=1, c=\wedge \}$, HTP $\rightarrow \{ i=2, c=\vee \}$. Total gains from profit shifting are obtained by substituting the optimal transfer price $p^{*}$ on the expected net profits $\mathbb{E}(\tilde{\Pi}(p^{*}))$ and comparing it with the net profits under the arm's length condition $\mathbb{E}(\tilde{\Pi}(\bar{p}^{c}))$. We find

\begin{equation} \label{profit shifting}
\mathbb{E}(\tilde{\Pi}(p^{*})) - \mathbb{E}(\tilde{\Pi}(\bar{p}^{c})) = (\tau_{2} - \tau_{1}) \cdot \Delta p^{*} m \cdot \dfrac{1}{1 + \gamma_{i}} > 0
\end{equation}

\noindent which is always positive for both LTP and HTP cases. 

\begin{proposition} \label{prop profit shifting gains}
Satisfying sufficient joint conditions for non-zero price gap, $\tau_{1} \neq \tau_{2}$, $z_{i} < \infty$ and $p^{c} \neq \bar{p}^{c}$, tax-induced variations in transfer prices always increase the expected net profits up to the optimal price gap
 \begin{equation*}
 \Delta p^{*}= \left( \dfrac{|\tau_{2} - \tau_{1}|}{(1 - e^{- \lambda_{i}}) \cdot (1+z_{i}) \cdot \tau_{i} \cdot \left( 1 + \frac{1}{\gamma_{i}} \right)} \right)^{\gamma_{i}} (p^{c} - \bar{p}^{c}) .
 \end{equation*}
  \begin{proof}
  Expected net profits with respect to the optimal transfer price $p^{*}$ and to the bound of the mode of the arm's length condition $\bar{p}^{c}$ are equal to
   \begin{equation*}
   \begin{array}{rcl}
   \mathbb{E}(\tilde{\Pi}(p^{*})) &=& (1 - \tau_{1})[R_{1}(s_{1}) - C_{1}(s_{1} + m)] + (1 - \tau_{2})[R_{2}(s_{2}) - C_{2}(s_{2} - m)] \\
   \\
   && + (\tau_{2} - \tau_{1}) \left[ \bar{p}^{c} + \left( \dfrac{|\tau_{2} - \tau_{1}|}{(1 - e^{- \lambda_{i}}) \cdot (1 + z_{i}) \cdot \tau_{i} \cdot \left( 1 + \frac{1}{\gamma_{i}} \right)} \right)^{\gamma_{i}} (p^{c} - \bar{p}^{c}) \right]m \\
   \\
   && - \left( \dfrac{\tau_{2} - \tau_{1}}{1 + \frac{1}{\gamma_{i}}} \right) \left( \dfrac{|\tau_{2} - \tau_{1}|}{(1 - e^{- \lambda_{i}}) \cdot (1 + z_{i}) \cdot \tau_{i} \cdot \left( 1 + \frac{1}{\gamma_{i}} \right)} \right)^{\gamma_{i}} (p^{c} - \bar{p}^{c})m ; \\
   \\
   \mathbb{E}(\tilde{\Pi}(\bar{p}^{c})) &=& (1 - \tau_{1})[R_{1}(s_{1}) - C_{1}(s_{1} + m)] + (1 - \tau_{2})[R_{2}(s_{2}) - C_{2}(s_{2} - m)] + (\tau_{2} - \tau_{1}) \bar{p}^{c} m . \\
   \end{array}
   \end{equation*}
  The difference $\mathbb{E}(\tilde{\Pi}(p^{*})) - \mathbb{E}(\tilde{\Pi}(\bar{p}^{c}))$ is equal to
   \begin{equation*}
   \begin{array}{l}
   \mathbb{E}(\tilde{\Pi}(p^{*})) - \mathbb{E}(\tilde{\Pi}(\bar{p}^{c})) = \\
   \\
   = (\tau_{2} - \tau_{1}) \left( \dfrac{1}{1 + \gamma_{i}} \right) \left( \dfrac{|\tau_{2} - \tau_{1}|}{(1 - e^{- \lambda_{i}}) \cdot (1 + z_{i}) \cdot \tau_{i} \cdot \left( 1 + \frac{1}{\gamma_{i}} \right)} \right)^{\gamma_{i}} (p^{c} - \bar{p}^{c})m \\
   \\ 
   = (\tau_{2} - \tau_{1}) \cdot \Delta p^{*} m \cdot \dfrac{1}{1 + \gamma_{i}} > 0
   \end{array}
   \end{equation*}
\noindent which is positive for both LTP and HTP cases derived in Eq. \ref{tp cases} as we have
   \begin{equation*}
   \begin{array}{lcl}
   \text{LTP} : \{ i=1, c = \wedge \} & \rightarrow & \{ \tau_{2} - \tau_{1} < 0, \Delta p^{*} < 0 \} ; \\
   \text{HTP} : \{ i=2, c = \vee \} & \rightarrow & \{ \tau_{2} - \tau_{1} > 0, \Delta p^{*} > 0 \} . \\ 
   \end{array} 
   \end{equation*}  
  \end{proof}
\end{proposition}

We state a relevant condition for the audit intensity $\lambda_{i} > 0$ derived from Eq. \ref{optimal p}:

\begin{proposition} \label{prop lambda condition}
  The optimal transfer price $p^{*}$ is a $\alpha$-cut of the fuzzy arm's length price $\tilde{P}^{c}$ only if the audit intensity $\lambda_{i}$ satisfies the condition
  \begin{equation*}
    \lambda_{i} \geq -\ln \left( 1 - \dfrac{|\tau_{2} - \tau_{1}|}{(1 + z_{i}) \cdot \tau_{i} \cdot \left( 1 + \frac{1}{\gamma_{i}} \right)} \right) .
  \end{equation*}
  \begin{proof}
    For the optimal transfer price $p^{*}$ to be an optimal $\alpha$-cut equal to $P_{= \alpha}^{c} = \{ p^{*} \in P : \mu_{\tilde{P}^{c}} (p^{*}) = \alpha \}$, specification in Eq. \ref{optimal p} requires the condition
    \begin{equation*}
      \left( \dfrac{|\tau_{2} - \tau_{1}|}{(1 - e^{- \lambda_{i}}) \cdot (1+z_{i}) \cdot \tau_{i} \cdot \left( 1 + \frac{1}{\gamma_{i}} \right)} \right)^{\gamma_{i}} \in [0,1] .
    \end{equation*}
    \noindent which implies $|\Delta p^{*}| \leq |p^{c} - \bar{p}^{c}|$. With respect to the domain of all variables within Eq. \ref{optimal p}, we observe that the only case where the condition $|\Delta p^{*}| \leq |p^{c} - \bar{p}^{c}|$ is violated is when the audit intensity $\lambda_{i}$ is sufficiently small, such that
    \begin{equation*}
      \{ \forall \delta > 0, \lambda_{i} > 0 : \lambda_{i} < \delta_{0} \} \rightarrow \left( \dfrac{|\tau_{2} - \tau_{1}|}{(1 - e^{- \lambda_{i}}) \cdot (1+z_{i}) \cdot \tau_{i} \cdot \left( 1 + \frac{1}{\gamma_{i}} \right)} \right)^{\gamma_{i}} > 1
    \end{equation*}
    \noindent for some value $\delta_{0}$. The necessary condition $|\Delta p^{*}| \leq |p^{c} - \bar{p}^{c}|$ implies
    \begin{equation*}
      \begin{array}{rcl}
        1 &\geq& \left( \dfrac{|\tau_{2} - \tau_{1}|}{(1 - e^{- \lambda_{i}}) \cdot (1+z_{i}) \cdot \tau_{i} \cdot \left( 1 + \frac{1}{\gamma_{i}} \right)} \right)^{\gamma_{i}} \\
        \\
        (1 - e^{- \lambda_{i}})^{\gamma_{i}} &\geq& \left( \dfrac{|\tau_{2} - \tau_{1}|}{(1+z_{i}) \cdot \tau_{i} \cdot \left( 1 + \frac{1}{\gamma_{i}} \right)} \right)^{\gamma_{i}} \\
        \\
        e^{- \lambda_{i}} &\leq& 1 - \dfrac{|\tau_{2} - \tau_{1}|}{(1+z_{i}) \cdot \tau_{i} \cdot \left( 1 + \frac{1}{\gamma_{i}} \right)} \\
        \\
        \lambda_{i} &\geq& - \ln \left( 1 - \dfrac{|\tau_{2} - \tau_{1}|}{(1+z_{i}) \cdot \tau_{i} \cdot \left( 1 + \frac{1}{\gamma_{i}} \right)} \right) . \\
      \end{array}
    \end{equation*}
    If this condition is not satisfied, the optimal transfer price $p^{*}$ as specified in Eq. \ref{optimal p} outbounds the interval of interest, $p^{*} \notin [p^{c}, \bar{p}^{c}]$, thus the solution of Eq. \ref{optimal p} is no more a $\alpha$-cut of the fuzzy arm's length price $\tilde{P}^{c}$.
  \end{proof}
\end{proposition}

\emph{Proposition \ref{prop lambda condition}} shows that the optimal transfer price as specified in Eq. \ref{optimal p} outbounds the interval of interest, $p^{*} \notin [p^{c}, \bar{p}^{c}]$ if the audit intensity is sufficiently weak. In this special case, the MNE can further increase the gains from profit shifting by disregarding the bounds of the fuzzy arm's length price $\tilde{P}^{c}$ when determining the transfer price $p$. It therefore implies:

\begin{corollary} \label{cor lambda condition}
  If the audit intensity $\lambda_{i} > 0$ does not satisfy the condition in \emph{Proposition \ref{prop lambda condition}}, maximisation object in Eq. \ref{max objective} has no general solution in terms of an optimal transfer price $p^{*}$.
  \begin{proof}
    Assume that the solution of Eq. \ref{optimal p} provides the inequality $|\Delta p^{*}| > |p^{c} - \bar{p}^{c}|$, so the condition in \emph{Proposition \ref{prop lambda condition}} is violated. Therefore, Eq. \ref{fuzzy arms length price}-\ref{intersection fuzzy arms length price} imply that the optimal transfer price $p^{*}$ has a membership grade equal to zero, $\mu_{\tilde{P}^{c}}(p^{*}) = 0$, so $p^{*}$ is considered a completely non-arm's length price. In this case, it is clear that the initial maximisation object in Eq. \ref{max objective} takes the form of a crisp linear function of $p$ in the first place, with no constraints, which is equal to
    \begin{equation*}
      \begin{array}{rcl}
        \smash{\underset{p \in P}{\max}} \quad \mathbb{E}(\Pi(\cdot)) &=& (1-\tau_{1})\pi_{1}+(1-\tau_{2})\pi_{2} \\
        &&- (1 - e^{- \lambda_{i}}) \cdot (1 + z_{i}) \cdot sgn(\tau_{2} - \tau_{1}) \tau_{i} \cdot (p - \bar{p}^{c})m \\
      \end{array}
    \end{equation*}
    \noindent with first derivative equal to
    \begin{equation*}
        \dfrac{\partial \mathbb{E}(\Pi(\cdot))}{\partial p} = (\tau_{2} - \tau_{1})m - (1 - e^{-\lambda_{i}}) \cdot (1 + z_{i}) \cdot sgn(\tau_{2} - \tau_{1}) \tau_{i} \cdot m = 0 
    \end{equation*}
    \noindent and second derivative equal to zero. The critical point at $\partial \mathbb{E}(\Pi(\cdot)) / \partial p = 0$ provides the same conditions as in Eq. \ref{tp cases}, for we have
    \begin{equation*}
    \begin{array}{lcl}
      \text{LTP} : \{ i = 1, \tau_{1} > \tau_{2} \} & \rightarrow & (\tau_{2} - \tau_{1}) = (1 - e^{-\lambda_{i}}) \cdot (1 + z_{i}) \cdot sgn(\tau_{2} - \tau_{1}) \tau_{i} < 0 ; \\
      \text{HTP} : \{ i = 2, \tau_{1} < \tau_{2} \} & \rightarrow & (\tau_{2} - \tau_{1}) = (1 - e^{-\lambda_{i}}) \cdot (1 + z_{i}) \cdot sgn(\tau_{2} - \tau_{1}) \tau_{i} > 0 . \\ 
    \end{array}	
    \end{equation*}
    Therefore, any changes in the transfer price $p$ towards the profit shifting direction increases the expected net profits with no upper bound, for both LTP ans HTP cases.
  \end{proof}
\end{corollary}

\emph{Corollary \ref{cor lambda condition}} simply shows that the MNE has full incentives to shift profits away from the high tax Country $i$ if the tax authority is lax. For a sufficiently weak audit intensity $\lambda_{i}$, the expected tax penalty becomes extremely low and linear with respect to the transfer price $p$ -- see \emph{Proposition \ref{prop lambda condition}}. In this case, it becomes profitable for the MNE to choose any tax-induced transfer price $p$, even if $p$ is considered a completely non-arm's length price. 

\subsection{Sensitivity Analyses} \label{Sensitivity Analyses}

Initially state the following:

\begin{corollary} \label{cor profit shifting intensif by m}
Under the optimality conditions regarding the price gap $\Delta p^{*}$, increase in intra-firm outputs $m$ always increases the total amount of profit shifting.
 \begin{proof}
 Derives directly from the gains of profit shifting in Eq. \ref{profit shifting}
  \begin{equation*}
  \dfrac{\partial [\mathbb{E}(\tilde{\Pi}(p^{*})) - \mathbb{E}(\tilde{\Pi}(\bar{p}^{c}))]}{\partial m} = (\tau_{2} - \tau_{1}) \cdot \Delta p^{*} \cdot \dfrac{1}{1 + \gamma_{i}} > 0 .
  \end{equation*}
 \end{proof}
\end{corollary}

Marginal changes in intra-firm output $m$ intensify the profit shifting amount, however intra-firm transfers depends on the product demand in Country 2, $s_{2}$. Assume that the demand in division 2 is not changed, but the MNE has flexibility to vary the application of internal output $m$ to provide revenues in division 2. Hence, the MNE can run a second optimisation stage to choose the optimal intra-firm output $m$. Differentiating the second-stage objective $\mathbb{E}(\tilde{\Pi}(p^{*}))$ with respect to intra-firm output $m$, subjected to the constraint $m \leq s_{2}$, we obtain

\begin{equation} \label{optimal m}
 \begin{array}{rcl}
 \dfrac{\partial \mathbb{E}(\tilde{\Pi}(p^{*},m))}{\partial m} &=& - (1 - \tau_{1}) \dfrac{\partial C_{1}(m)}{\partial m} + (1 - \tau_{2}) \dfrac{\partial C_{2}(m)}{\partial m} + (\tau_{2} - \tau_{1}) \left( \bar{p}^{c} + \Delta p^{*} \dfrac{1}{1 + \gamma_{i}} \right)  \\
 \\
 && - L(s_{2} - m) = 0 ; \\
 \\
 (1 - \tau_{1}) \dfrac{\partial C_{1}(m)}{\partial m} &=& (1 - \tau_{2}) \dfrac{\partial C_{2}(m)}{\partial m} + (\tau_{2} - \tau_{1}) \left( \bar{p}^{c} + \Delta p^{*} \dfrac{1}{1 + \gamma_{i}} \right) - L(s_{2} - m) \\
 \end{array}
\end{equation}

\noindent with conditions $- \partial C_{1}(m) / \partial m < 0$, $\partial C_{2}(m) / \partial m > 0$, $L(s_{2} - m) = 0$, where $L(\cdot)$ is the Lagrangian multiplier function. If the constraint is not binding, $m < s_{2}$, we have $L(\cdot) = 0$. In Eq. \ref{optimal m}, the exogenous effect of the intra-firm transaction $(\tau_{2} - \tau_{1}) \bar{p}^{c}$ follows the direction of the profit shifting incentive -- LTP implies $(\tau_{2} - \tau_{1}) \bar{p}^{\wedge} < 0$ and HTP implies $(\tau_{2} - \tau_{1}) \bar{p}^{\vee} > 0$, $\bar{p}^{c} \in P \in \mathbb{R}_{+}$, $c = \{ \wedge,\vee \}$. However, the effect of the optimal price gap is always positive, $(\tau_{2} - \tau_{1}) \Delta p^{*} \cdot [1/(1 + \gamma_{i})] > 0$, $\gamma_{i} \in (0,1]$ -- see the \emph{Corollary \ref{cor profit shifting intensif by m}} . Equality of net marginal costs occurs only in the extreme LTP case where the effect of the optimal price gap completely neutralises the exogenous arm's length effect, i.e. iff $\bar{p}^{c} = \Delta p^{*} \cdot [1/(1 + \gamma_{i})]$. Hence, at the optimal intra-firm output $m^{*}$, the generalised condition $(1 - \tau_{1}) \cdot [\partial C_{1}(m)/\partial m] \neq (1 - \tau_{2}) \cdot [\partial C_{2}(m)/\partial m]$ is attained in most of cases.

In special, Eq. \ref{optimal m} indicates that different tax rates disturb the effect of the $m$-elasticity of substitution between costs $C_{1}(m)$ and $C_{2}(m)$ over the expected net profits $\mathbb{E}(\tilde{\Pi}(p,m))$. The effect is equal to $\frac{1 - \tau_{1}}{1 - \tau_{2}} \cdot \varepsilon_{C_{1},C_{2}}$, $\tau_{i} \in [0,1]$, where $\varepsilon_{C_{1},C_{2}}$ is the $m$-elasticity of substitution\footnote{Formally, the effect of the $m$-elasticity is equal to
\begin{equation*}
\dfrac{1 - \tau_{1}}{1 - \tau_{2}} \cdot \varepsilon_{C_{1},C_{2}} - \left( \dfrac{\tau_{2} - \tau_{1}}{1 - \tau_{2}} \right) \left( \dfrac{\bar{p}^{c} m}{C_{1}(m)} \right) \cdot \varepsilon_{\bar{p}^{c},C_{2}} -\left( \dfrac{\tau_{2} - \tau_{1}}{1 - \tau_{2}} \right) \left( \dfrac{1}{1 + \gamma_{i}} \right) \left( \dfrac{\Delta p^{*} m}{C_{1}(m)} \right) \cdot \varepsilon_{\Delta p^{*},C_{2}}
\end{equation*}
\noindent where $\varepsilon_{\bar{p}^{c},C_{2}}$ is the $m$-elasticity between the arm's length parameter $\bar{p}^{c}$ and $C_{2}(m)$, and $\varepsilon_{\Delta p^{*},C_{2}}$ is the $m$-elasticity between $\Delta p^{*}$ and $C_{2}(m)$. Eq. \ref{optimal p} implies both $\partial \bar{p}^{c}/ \partial m = 0 \rightarrow \partial \bar{p}^{c}/ \partial C_{2}(m) = 0$, $\partial \Delta p^{*}/ \partial m = 0 \rightarrow \partial \Delta p^{*}/ \partial C_{2}(m) = 0$, therefore we have $\varepsilon_{\bar{p}^{c},C_{2}} = 0$, $\varepsilon_{\Delta p^{*},C_{2}} = 0$.} between $C_{1}(m)$ and $C_{2}(m)$. 

With respect to the profit shifting incentive $\tau_{2} - \tau_{1}$, $\tau_{1} \neq \tau_{2}$, it is clear that the optimal transfer price $p^{*}$ is affected by changes in tax rate $\tau_{i} \in [0,1]$. Differentiating Eq. \ref{optimal p} with respect to $\tau_{i}$, we obtain the following standard-form equations\footnote{Derivation of Eq. \ref{optimal dp dt}-\ref{optimal ddp dtt} in Appendix.}:

\begin{equation} \label{optimal dp dt}
 \dfrac{\partial p^{*}}{\partial \tau_{i}} = \dfrac{\gamma_{i} (p^{c} - \bar{p}^{c})}{\left( (1 - e^{-\lambda_{i}}) \cdot (1 + z_{i}) \cdot \left(1 + \frac{1}{\gamma_{i}} \right) \right)^{\gamma_{i}}} \cdot \dfrac{\tau_{i} - |\tau_{2} - \tau_{1}|}{\tau_{i}^{1 + \gamma_{i}} \cdot |\tau_{2} - \tau_{1}|^{1 - \gamma_{i}}} ;
\end{equation}
\\
\begin{equation} \label{optimal ddp dtt}
 \dfrac{\partial^{2} p^{*}}{\partial \tau_{i}^{2}} = \dfrac{\partial p^{*}}{\partial \tau_{i}} \cdot \dfrac{(\gamma_{i} - 1) \cdot \tau_{i} - (\gamma_{i} + 1) \cdot |\tau_{2} - \tau_{1}|}{\tau_{i} \cdot |\tau_{2} - \tau_{1}|} .
\end{equation}

First, Eq. \ref{optimal dp dt} shows that the variation $\partial p^{*} / \partial \tau_{i}$ follows the profit shifting direction.


\begin{proposition} \label{proposition optimal dp dt}
  Variation in the optimal transfer price $p^{*}$ with respect to marginal changes in the tax rate $\tau_{i}$ of the harmed Country $i$ follows the direction of the profit shifting incentive $\tau_{2} - \tau_{1}$, such that $sgn(\partial p^{*} / \partial \tau_{i}) = sgn(\Delta p^{*})$. Variation $\partial p^{*} / \partial \tau_{i}$ is equal zero iff $\tau_{j \neq i} = 0$.
  \begin{proof}
    From Eq. \ref{optimal dp dt}, the difference $\tau_{i} - |\tau_{2} - \tau_{1}| > 0$, $\tau_{i} \in [0,1]$, $\tau_{j \neq i} \neq 0$, $i,j = \{ 1,2 \}$ is always positive. In this case, the following conditions are satisfied for each LTP and HTP cases:
    \begin{equation*}
      \begin{array}{lcl}
        \text{LTP} : \{ i=1, c = \wedge \} & \rightarrow & \{ \Delta p^{*} < 0,  p^{c} - \bar{p}^{c} < 0 \rightarrow \partial p^{*} / \partial \tau_{i} < 0 \} ; \\
        \text{HTP} : \{ i=2, c = \vee \} & \rightarrow & \{ \Delta p^{*} > 0,  p^{c} - \bar{p}^{c} > 0 \rightarrow \partial p^{*} / \partial \tau_{i} > 0 \} . \\ 
      \end{array} 
    \end{equation*}
    For the limiting case where $\tau_{j \neq i} = 0$, Eq. \ref{optimal p} is no more a function of $\tau_{i}$, so we clearly have $\tau_{j \neq i} = 0 \rightarrow \partial p^{*}/\partial \tau_{i} = 0$. 
  \end{proof}
\end{proposition}

As it is intuitively conjectured, \emph{Proposition \ref{proposition optimal dp dt}} shows that marginal increments in the tax rate $\tau_{i}$ of the harmed Country $i$ widens the optimal price gap $\Delta p^{*}$ thus to shift more profits away from Country $i$, since it represents an increase in the profit shifting incentive. Reductions in tax rate $\tau_{i}$ cause the reverse effect. On the other hand, for marginal changes in tax rate $\tau_{j \neq i}$, $i,j = \{1,2 \} $ of the non-harmed Country $j$, variation $\partial p^{*} / \partial \tau_{j} $ takes the opposite direction, e.g. a marginal increase in $\tau_{j \neq i}$ shrinks the optimal price gap $\Delta p^{*}$ and reduces the gains from profit shifting\footnote{For changes in $\tau_{j \neq i}$, $i,j = \{1,2 \} $, we have the condition
\begin{equation*}
  sgn \left( \dfrac{\partial p^{*}}{\partial \tau_{i}} \right) = - sgn \left( \dfrac{\partial p^{*}}{\partial \tau_{j}} \right) \rightarrow sgn \left( \dfrac{\partial p^{*}}{\partial \tau_{j}} \right) = - sgn(\Delta p^{*}) .
\end{equation*}
}. For the limiting case where the tax rate of the non-harmed Country $j$ is zero, $\tau_{j \neq i} = 0$, changes in the tax rate $\tau_{i}$ do not affect the optimal price gap $\Delta p^{*}$. The general condition for Eq. \ref{optimal dp dt} to be linear occurs at the arm's length tolerance parameter equal to

\begin{equation*}
 \gamma_{i}(\tau_{i},|\tau_{2} - \tau_{1}|) = \dfrac{W[(\tau_{i} \cdot |\tau_{2} - \tau_{1}|) \cdot ( \ln (|\tau_{2} - \tau_{1}|) - \ln (\tau_{i}))]}{ \ln (|\tau_{2} - \tau_{1}|) - \ln (\tau_{i})} \in (0,1]
\end{equation*}

\noindent with $\tau_{j \neq i} \neq 0$, where $W(\tau_{i}, |\tau_{2} - \tau_{1}|)$ is the Lambert product log function\footnote{Lambert product log function is defined as the following: for an exponential function $x e^x$, the inverse function is such as it satisfies the condition $x = f^{-1}(x e^x) = W(x e^x)$, where $W(x)$ is called the Lambert product log function. It is applied for a general power function $f(x) = x a^x$, $a \in \mathbb{R}$ as $x = \frac{W(f(x) \cdot \ln (a))}{ \ln (a)}$.}. 

Furthermore, Eq. \ref{optimal ddp dtt} describes the slope of changes in the optimal transfer price $p^{*}$ as the tax rate $\tau_{i}$ changes. Under the scope of both LTP and HTP cases, Eq. \ref{optimal ddp dtt} shows that the slope of the variation $\partial p^{*} / \partial \tau_{i}$ is opposite to the profit shifting direction; the slope of $\partial p^{*} / \partial \tau_{i}$ is strictly increasing for the LTP case and strictly decreasing for the HTP case. 


\begin{proposition} \label{proposition optimal ddp dtt}
  The slope of the variation $\partial p^{*} / \partial \tau_{i}$ is opposite to the profit shifting direction, such that $sgn(\partial^{2} p^{*}/\partial \tau_{i}^{2}) = -sgn(\partial p^{*} / \partial \tau_{i})$. 
  \begin{proof}
    First, the multiplier at the right hand side of Eq. \ref{optimal ddp dtt} equal to
    \begin{equation*}
      \dfrac{\varepsilon_{p^{*},\tau_{i}}^{2}}{\tau_{i}} = \dfrac{(\gamma_{i} - 1) \cdot \tau_{i} - (\gamma_{i} + 1) \cdot |\tau_{2} - \tau_{1}|}{\tau_{i} \cdot |\tau_{2} - \tau_{1}|}
    \end{equation*}
    \noindent is defined as the second-order $\tau_{i}$-semi-elasticity of the optimal transfer price $p^{*}$, thus Eq. \ref{optimal ddp dtt} is equal to Eq. \ref{optimal dp dt} multiplied by \smash{$\varepsilon_{p^{*},\tau_{i}}^{2} / \tau_{i}$}. We notice that \smash{$\varepsilon_{p^{*},\tau_{i}}^{2} / \tau_{i} < 0$}, $\tau_{i} \in [0,1]$, $\gamma_{i} \in (0,1]$ is always negative for both LTP and HTP cases. \emph{Proposition \ref{proposition optimal dp dt}} shows that $sgn(\partial p^{*} / \partial \tau_{i}) = sgn(\Delta p^{*})$, therefore it implies
    \begin{equation*}
      sgn \left( \dfrac{\partial^{2} p^{*}}{\partial \tau_{i}^{2}} \right) = -sgn \left( \dfrac{\partial p^{*}}{\partial \tau_{i}} \right) = -sgn (\Delta p^{*}) .
    \end{equation*}
    If $\tau_{i} \rightarrow \tau_{j}$, $i,j = \{ 1,2 \}$, the tax differential tends to zero, $\tau_{2} - \tau_{1} \rightarrow 0$, so Eq. \ref{optimal ddp dtt} diverges for any $\gamma_{i} < 1$ as follows:
    \begin{equation*}
      \gamma_{i} < 1 \rightarrow \displaystyle\lim_{\tau_{i} \rightarrow \tau_{j}} \dfrac{\gamma_{i} (p^{c} - \bar{p}^{c})}{\left( (1 - e^{-\lambda_{i}}) \cdot (1 + z_{i}) \cdot \left(1 + \frac{1}{\gamma_{i}} \right) \right)^{\gamma_{i}}} \cdot \dfrac{(\gamma_{i} - 1)}{\tau_{i}^{\gamma_{i}} \cdot \underset{\rightarrow 0_{+}}{|\tau_{2} - \tau_{1}|}^{2 - \gamma_{i}} } = -sgn(\tau_{2} - \tau_{1}) \cdot \infty
    \end{equation*}
    \noindent where $sgn(\tau_{2} - \tau_{1}) = sgn(p^{c} - \bar{p}^{c})$, which implies
    \begin{equation*}
      sgn \left( \dfrac{\partial^{2} p^{*}}{\partial \tau_{i}^{2}} \right) = -sgn(\tau_{2} - \tau_{1}) = -sgn \left( \dfrac{\partial p^{*}}{\partial \tau_{i}} \right) = -sgn (\Delta p^{*}) .
    \end{equation*}
    The slope of $\partial p^{*}/\partial \tau_{i}$ is equal to zero only in the special case where $\gamma_{i} = 1$, $\tau_{i} = \tau_{j}$, i.e. it does not fit either LTP or HTP cases. Therefore, both LTP and HTP cases strictly satisfy the condition $sgn(\partial^{2} p^{*}/\partial \tau_{i}^{2}) = -sgn(\partial p^{*} / \partial \tau_{i})$.
  \end{proof}
\end{proposition}

Eq. \ref{optimal ddp dtt} shows that the variation $\partial p^{*} / \partial \tau_{i}$ is elastic if the tax differential $\tau_{2} - \tau_{1}$ is narrow, since it implies that the initial shifting incentive is weak and the maximising price gap $\Delta p^{*}$ is rather small -- see Eq. \ref{optimal p}. In this case, small marginal changes in $\tau_{i}$ produce large marginal impacts on the optimal transfer price $p^{*}$. On the other hand, a large profit shifting incentive $\tau_{2} - \tau_{1}$ implies that the initial price gap $\Delta p^{*}$ is already wide, and further marginal changes in the tax rate $\tau_{i}$ produce a weaker impact. Variation $\partial p^{*} / \partial \tau_{i}$ becomes inelastic as the tax rate approaches the boundary $\tau_{i} \rightarrow 1$, $|\tau_{2} - \tau_{1}| < 1$.

\begin{corollary} \label{corollary unitary tau elasticity}
  Unitary $\tau_{i}$-elasticity of the optimal price gap $\Delta p^{*}$ occurs at the equality
  \begin{equation*}
    \gamma_{i}(\tau_{i} - |\tau_{2} - \tau_{1}|) = |\tau_{2} - \tau_{1}| 
  \end{equation*}
  \noindent such that the arm's length tolerance parameter $\gamma_{i}$ scales the impact of the difference $\tau_{i} - |\tau_{2} - \tau_{1}|$ within Eq. \ref{optimal dp dt}.
  \begin{proof}
    Derives directly from Eq. \ref{optimal p}:
    \begin{equation*}
      \begin{array}{rcl}
        \varepsilon_{\Delta p^{*}, \tau_{i}} = \dfrac{\partial \Delta p^{*}}{\partial \tau_{i}} \cdot \dfrac{\tau_{i}}{\Delta p^{*}} &=& \gamma_{i} \cdot\dfrac{\tau_{i} - |\tau_{2} - \tau_{1}|}{|\tau_{2} - \tau_{1}|} ; \\
        \\
        \varepsilon_{\Delta p^{*}, \tau_{i}} = 1 & \rightarrow & \gamma_{i}(\tau_{i} - |\tau_{2} - \tau_{1}|) = |\tau_{2} - \tau_{1}| . \\
      \end{array}
      \end{equation*}
  \end{proof}
\end{corollary}

In overall, Eq. \ref{optimal dp dt}-\ref{optimal ddp dtt} show that the variation $\partial p^{*}/\partial \tau_{i}$ converges if the general conditions $\tau_{i} > 0$, $\tau_{i} \neq \tau_{j}$ are satisfied. The general convergence at individually increasing tax rates $\tau_{i} \rightarrow 1$, $\tau_{j \neq i} \rightarrow 1$ are obtained as follows:

\begin{equation*}
  \lim_{\tau_{i} \to 1} \dfrac{\partial p^{*}}{\partial \tau_{i}} = \dfrac{\gamma_{i} (p^{c} - \bar{p}^{c})}{\left( (1 - e^{-\lambda_{i}}) \cdot (1 + z_{i}) \cdot \left(1 + \frac{1}{\gamma_{i}} \right) \right)^{\gamma_{i}}} \cdot \dfrac{\tau_{j}}{|1 - \tau_{j}|^{1 - \gamma_{i}}} ;
\end{equation*}
\begin{equation*}
  \lim_{\tau_{j \neq i} \to 1} \dfrac{\partial p^{*}}{\partial \tau_{i}} = \dfrac{\gamma_{i} (p^{c} - \bar{p}^{c})}{\left( (1 - e^{-\lambda_{i}}) \cdot (1 + z_{i}) \cdot \left(1 + \frac{1}{\gamma_{i}} \right) \right)^{\gamma_{i}}} \cdot \dfrac{2 \tau_{i} - 1}{\tau_{i}^{1 + \gamma_{i}} \cdot |\tau_{i} - 1|^{1 - \gamma_{i}}} .
\end{equation*}

We also have $\lim_{\gamma_{i} \to 0} \partial p^{*} / \partial \tau_{i} = 0$. Nonetheless, in the limiting case of the tightest tax assessment $\gamma_{i} \rightarrow 1$, we obtain a special convergence for $\tau_{i} \rightarrow \tau_{j}$ equal to

\begin{equation*}
  \underset{\substack{\tau_{i} \to \tau_{j} \\ \gamma_{i} \to 1 }}{\lim} \dfrac{\partial p^{*}}{\partial \tau_{i}} = \dfrac{(p^{c} - \bar{p}^{c})}{ 2 \tau_{i} (1 - e^{-\lambda_{i}}) (1 + z_{i})} ,
\end{equation*}

\noindent from which we obtain

\begin{equation*}
  \underset{\substack{\tau_{j \neq i} \to 1 \\ \gamma_{i} \to 1 }}{\lim} \left( \lim_{\tau_{i} \rightarrow 1} \dfrac{\partial p^{*}}{\partial \tau_{i}} \right) = \underset{\substack{\tau_{i} \to 1 \\ \gamma_{i} \to 1 }}{\lim} \left( \lim_{\tau_{j \neq i} \rightarrow 1} \dfrac{\partial p^{*}}{\partial \tau_{i}} \right) = \lim_{\tau_{i} \to 1} \left( \underset{\substack{\tau_{i} \to \tau_{j} \\ \gamma_{i} \to 1 }}{\lim} \dfrac{\partial p^{*}}{\partial \tau_{i}} \right) = \dfrac{(p^{c} - \bar{p}^{c})}{ 2 (1 - e^{-\lambda_{i}}) (1 + z_{i})} .
\end{equation*}

Otherwise, variation $\partial p^{*}/\partial \tau_{i}$ diverges as follows:

\begin{equation*}
 \lim_{\tau_{i} \to \tau_{j}} \dfrac{\partial p^{*}}{\partial \tau_{i}} = sgn(\tau_{2} - \tau_{1}) \cdot \infty, \quad \gamma_{i} < 1 ; 
\end{equation*}
\begin{equation*}
 \lim_{\tau_{i} \to 0} \dfrac{\partial p^{*}}{\partial \tau_{i}} = -sgn(\tau_{2} - \tau_{1}) \cdot \infty, \quad \tau_{j \neq i} > 0 .
\end{equation*}

With respect to the other country-level variables related with the audit intensity $\lambda_{i} > 0$, tax penalty $z_{i} > 0$ and the arm's length tolerance parameter $\gamma_{i} \in (0,1]$, specification of the optimal price gap $\Delta p^{*}$ in Eq. \ref{optimal p} shows an elaborate effect. Variation of the optimal price gap $\Delta p^{*}$ with respect to each individual country-level variable is equal to:


\begin{equation} \label{optimal dp dlambda}
  \dfrac{\partial p^{*}}{\partial \lambda_{i}} = \Delta p^{*} \cdot \dfrac{-\gamma_{i}e^{-\lambda_{i}}}{1 - e^{-\lambda_{i}}} ;
\end{equation}
\begin{equation} \label{optimal dp dz}
  \dfrac{\partial p^{*}}{\partial z_{i}} = \Delta p^{*} \cdot \dfrac{- \gamma_{i}}{1 + z_{i}} ; 
\end{equation}
\begin{equation} \label{optimal dp dgamma}
  \dfrac{\partial p^{*}}{\partial \gamma_{i}} = \Delta p^{*} \cdot \left[ \ln \left( \dfrac{|\tau_{2} - \tau_{1}|}{(1 - e^{-\lambda_{i}}) \cdot (1 + z_{i}) \cdot\tau_{i} \cdot \left( 1 + \frac{1}{\gamma_{i}} \right)} \right) + \dfrac{1}{1 + \gamma_{i}} \right] . 
\end{equation}

First regarding Eq. \ref{optimal dp dlambda}-\ref{optimal dp dz}, both clearly present a negative effect on the optimal price gap $\Delta p^{*}$ for the complete interval of interest $p^{*} \in [p^{c},\bar{p}^{c}]$, which is consistent with the intuitive premise, e.g. a marginal increase in the audit intensity $\lambda_{i}$ or in the penalty rate $z_{i}$ increases the expected tax penalty $\mathbb{E}(\tilde{Z}_{i}(\cdot))$, thus it shortens the optimal price gap $\Delta p^{*}$; a decrease in $\lambda_{i}$ or $z_{i}$ produces the opposite effect. 

For the Eq. \ref{optimal dp dgamma}, on the other hand, the effect is not necessarily negative on the full interval $[p^{c},\bar{p}^{c}]$, for all $\forall \gamma_{i} \in (0,1]$. In the general case, the variation $\partial p^{*} / \partial \gamma_{i}$ represents a non-positive effect over the optimal transfer price $p^{*}$ iff it satisfies the inequality

\begin{equation*}
  \ln \left( \dfrac{|\tau_{2} - \tau_{1}|}{(1 - e^{-\lambda_{i}}) \cdot (1 + z_{i}) \cdot\tau_{i} \cdot \left( 1 + \frac{1}{\gamma_{i}} \right)} \right) \leq - \dfrac{1}{1 + \gamma_{i}} < 0 ;
\end{equation*}

\noindent otherwise, the effect follows the profit shifting direction. This positive effect is counter-intuitive at first, for we expect that changes in the arm's length tolerance parameter $\gamma_{i} \in (0,1]$ to produce only effects that are opposite to the shifting incentive. Nonetheless, notice that the specification of the fuzzy arm's length price $\tilde{P}^{c}$ in terms of $p$ produces a marginal adjustment effect on the gains from profit shifting equal to $1/(1 + \gamma_{i})$ -- see Eq. \ref{optimal p}. It means that the effect of marginal changes in the tolerance parameter $\gamma_{i}$ must be negative and must outburst the marginal effect of $1/(1 + \gamma_{i})$, in order to produce a negative effect on $p^{*}$. This last outcome\footnote{More specifically, marginal changes in the tolerance parameter $\gamma_{i}$ produce a positive effect on $p^{*}$ iff Eq. \ref{optimal p} implies the inequality $|\tau_{2} - \tau_{1}| > (1 - e^{-\lambda_{i}}) \cdot (1 + z_{i}) \cdot \tau_{i} \cdot ( 1 + 1/\gamma_{i} )$. In this case, the optimal transfer price $p^{*}$ is increasing at $\gamma_{i}$, although it implies beforehand that $p^{*}$ is not a $\alpha$-cut of the fuzzy arm's length price $\tilde{P}^{c}$ -- see \emph{Corollary \ref{cor lambda condition}}.} is specially due to the specification of $\tilde{P}^{c}$ in terms of $p$ within Eq. \ref{max objective parametric in p}.

\subsection{Modelling a General Tax Enforcement Effect} \label{Modelling a General Tax Enforcement Effect}

While Eq. \ref{optimal dp dlambda}-\ref{optimal dp dgamma} show how marginal changes in individual anti-shifting variables affect the optimal transfer price $p^{*}$, the influence of a general enforcing behaviour from the harmed Country $i$ may be reflected simultaneously in more than one variable. In special, it is safe to assume that the audit intensity $\lambda_{i} > 0$ and the arm's length tolerance parameter $\gamma_{i} \in (0,1]$ are both related with some common measure of tax enforcement applied by Country $i$. 


Let $\dot{\gamma_{i}} \in (0,1]$ be a regularised tax enforcement measure for the Country $i$, such that the arm's length tolerance parameter $\gamma_{i}$ is a monotone order-preserving function $\gamma_{i} = g(\dot{\gamma_{i}}) : (0,1] \rightarrow (0,1]$, for a weak tax enforcement implies $\dot{\gamma_{i}} \rightarrow 0$, while strong tax enforcement implies $\dot{\gamma_{i}} \rightarrow 1$. Moreover, assume that the tax audit intensity $\lambda_{i} > 0$ varies with respect to tax enforcement $\dot{\gamma_{i}}$, thus the audit intensity becomes a non-homogeneous Poisson rate function with respect to the tax enforcement, $\lambda_{i}(\dot{\gamma_{i}}) : (0,1] \rightarrow \mathbb{R}_{++}$. Assume that $\lambda_{i}(\dot{\gamma_{i}})$ is continuous. Hence, if the number of tax audits performed by Country $i$ is $q(\dot{\gamma_{i}}) \in \mathbb{N}$ and the tax enforcement level $\dot{\gamma_{i}} \in (0,1]$ implies $q(\dot{\gamma_{i}}) \geq 0$, $\lim_{\dot{\gamma_{i}} \rightarrow 0} q(\dot{\gamma_{i}}) = 0$, then the probability of exact $q(\dot{\gamma_{i}}) = k$ audits is

\begin{equation*}
  \mathbb{P}(q(\dot{\gamma_{i}}) = k, \lambda_{i}(\dot{\gamma_{i}})) = \dfrac{(\Lambda_{i}(0, \dot{\gamma_{i}}))^{k} \cdot e^{-\Lambda_{i}(0, \dot{\gamma_{i}})}}{k !} 
\end{equation*}

\noindent with the non-homogeneous Poisson intensity parameter equal to

\begin{equation} \label{nonhom audit intensity}
  \Lambda_{i}(0, \dot{\gamma_{i}}) = \int_{\rightarrow 0_{+}}^{\dot{\gamma_{i}}} \lambda_{i}(y) \, \mathrm{d}y .
\end{equation}

Eq. \ref{nonhom audit intensity} derives from the non-homogeneous Poisson condition 

\begin{equation*}
   \mathbb{P}[q(\dot{\gamma_{i}} + \Delta \dot{\gamma_{i}}) - q(\dot{\gamma_{i}}) = 1, \lambda_{i}(\dot{\gamma_{i}})] \approx \lambda_{i}(\dot{\gamma_{i}}) \Delta \dot{\gamma_{i}} 
\end{equation*}

\noindent which says that the incremental probability of one additional tax audit by Country $i$ is approximate to a linear relation between the rate function $\lambda_{i}(\dot{\gamma_{i}})$ at $\dot{\gamma_{i}}$ and the variation in tax enforcement $\Delta \dot{\gamma_{i}}$. The total probability of tax audit for the MNE derives directly from Eq. \ref{total prob tax audit} and Eq. \ref{nonhom audit intensity} and is equal to 

\begin{equation*}
  \mathbb{P}(q(\dot{\gamma_{i}}) > 0, \lambda_{i}(\dot{\gamma_{i}})) = 1 - \mathbb{P}(q(\dot{\gamma_{i}}) = 0, \lambda_{i}(\dot{\gamma_{i}})) = 1 - e^{-\Lambda_{i}(0,\dot{\gamma_{i}})} .
\end{equation*}

Specification of the audit rate function $\lambda_{i}(\dot{\gamma_{i}})$ is not a straight task. In general, existing studies suggest that higher tax enforcement implies in more frequent audits, although the increments on the number of tax audits vary through the enforcement range $\dot{\gamma_{i}} \in (0,1]$ \cite{alm2012}, so we assume that the audit rate is clearly non-decreasing as the tax enforcement $\dot{\gamma_{i}}$ increases.

To simplify the analysis, define a general function $f(\dot{\gamma_{i}}) : (0,1] \rightarrow \mathbb{R}_{++}$, $f(\dot{\gamma_{i}})$ is monotone continuous for $ \forall \dot{\gamma_{i}} \in (0,1)$, and it satisfies $\lim_{\dot{\gamma_{i}} \to 0} f(\dot{\gamma_{i}}) = 0$, $\lim_{\dot{\gamma_{i}} \to 1} f(\dot{\gamma_{i}}) = 1$. Under adequate conditions\footnote{Derivation of the conditions for Eq. \ref{prob tax audit parametric in enforcement} in Appendix.} regarding the audit rate function $\lambda_{i}(\dot{\gamma_{i}})$ and the tax enforcement $\dot{\gamma_{i}}$, we argue that the the audit probability can be parametrised with respect to the variable $\dot{\gamma_{i}}$ as

\begin{equation} \label{prob tax audit parametric in enforcement}
 \mathbb{P}(q(\dot{\gamma_{i}}) > 0, \lambda_{i}(\dot{\gamma_{i}})) = f(\dot{\gamma_{i}}) 
\end{equation}

\noindent so we are able to define the optimal transfer price in Eq. \ref{optimal p} in terms of the tax enforcement variable $\dot{\gamma_{i}} \in (0,1]$, equal to

\begin{equation*}
  p^{*} (\dot{\gamma_{i}}, \cdot) = \bar{p}^{c} + \left( \dfrac{|\tau_{2} - \tau_{1}|}{f(\dot{\gamma_{i}}) \cdot (1+z_{i}) \cdot \tau_{i} \cdot \left( 1 + \frac{1}{g(\dot{\gamma_{i}})} \right)} \right)^{g(\dot{\gamma_{i}})} (p^{c} - \bar{p}^{c}) .
\end{equation*}

From now on, simplify the notation of both general functions as $g(\dot{\gamma_{i}}) = g_{\dot{\gamma_{i}}}$ and $f(\dot{\gamma_{i}}) = f_{\dot{\gamma_{i}}}$. On the bounded domain $\dot{\gamma_{i}} \in (0,1]$, both functions $g_{\dot{\gamma_{i}}}$, $f_{\dot{\gamma_{i}}}$ have the same limiting values on the boundaries, $\dot{\gamma_{i}} \rightarrow 0$ and $\dot{\gamma_{i}} \rightarrow 1$, which are equal to

\begin{equation*}
  \begin{array}{rcccl}
    \displaystyle\lim_{\dot{\gamma_{i}} \to 0} g_{\dot{\gamma_{i}}} &=& \displaystyle\lim_{\dot{\gamma_{i}} \to 0} f_{\dot{\gamma_{i}}} &=& 0 ; \\
    \\
    \displaystyle\lim_{\dot{\gamma_{i}} \to 1} g_{\dot{\gamma_{i}}} &=& \displaystyle\lim_{\dot{\gamma_{i}} \to 1} f_{\dot{\gamma_{i}}} &=& 1  \\
  \end{array}
\end{equation*}

\noindent by definition, regardless of their slopes. Since functions $f_{\dot{\gamma_{i}}}$, $g_{\dot{\gamma_{i}}}$ are continuous on the complete domain, $\forall \dot{\gamma_{i}} \in (0,1]$, we observe that any monotonic map $F(f(\cdot)) : (0,1] \rightarrow (0,1]$ implies $|F(f_{\dot{\gamma_{i}}}) - F(g_{\dot{\gamma_{i}}})| < \infty $. Therefore, it implies the following:


\begin{proposition} \label{proposition lim optimal p in enforcement}
  For the optimal transfer price $p^{*}(\dot{\gamma_{i}}, \cdot)$ parametrised with respect to the tax enforcement variable $\dot{\gamma_{i}} \in (0,1]$, the maximising prices at the boundaries of the domain, $\dot{\gamma_{i}} \rightarrow 0$ and $\dot{\gamma_{i}} \rightarrow 1$ are equal to
  
  \begin{equation*}
    \begin{array}{rcl}
      \displaystyle\lim_{\dot{\gamma_{i}} \to 0} p^{*}(\dot{\gamma_{i}}, \cdot) &=& p^{c} ; \\
      \\
      \displaystyle\lim_{\dot{\gamma_{i}} \to 1} p^{*}(\dot{\gamma_{i}}, \cdot) &=& \bar{p}^{c} + \dfrac{|\tau_{2} - \tau_{1}|}{2 (1+z_{i}) \tau_{i}} \cdot (p^{c} - \bar{p}^{c}) . \\
    \end{array}
\end{equation*}

  \begin{proof}
    For any marginal change in the tax enforcement $\dot{\gamma_{i}}$, variation in $p^{*}(\dot{\gamma_{i}}, \cdot)$ depends on the effect of both $g_{\dot{\gamma_{i}}}$, $f_{\dot{\gamma_{i}}}$, which is equal to $[g_{\dot{\gamma_{i}}} / (f_{\dot{\gamma_{i}}}(1 + g_{\dot{\gamma_{i}}}))]^{g_{\dot{\gamma_{i}}}}$. For the upper bound $\dot{\gamma_{i}} \rightarrow 1$, we clearly have
    
    \begin{equation*}
      \begin{array}{rcl}
        \displaystyle\lim_{\dot{\gamma_{i}} \rightarrow 1} \left( \dfrac{g_{\dot{\gamma_{i}}}}{f_{\dot{\gamma_{i}}}(1 + g_{\dot{\gamma_{i}}})} \right)^{g_{\dot{\gamma_{i}}}} = \dfrac{1}{2} & \rightarrow & \displaystyle\lim_{\dot{\gamma_{i}} \to 1} p^{*}(\dot{\gamma_{i}}, \cdot) = \bar{p}^{c} + \dfrac{|\tau_{2} - \tau_{1}|}{2 (1+z_{i}) \tau_{i}} \cdot (p^{c} - \bar{p}^{c}) . \\
      \end{array}
    \end{equation*}
    
    For the lower bound $\dot{\gamma_{i}} \rightarrow 0$, we derive the following:
    
    \begin{equation*}
      \begin{array}{rcl}
        \ln \left( \displaystyle\lim_{\dot{\gamma_{i}} \rightarrow 0} \left( \dfrac{g_{\dot{\gamma_{i}}}}{f_{\dot{\gamma_{i}}}(1 + g_{\dot{\gamma_{i}}})} \right)^{g_{\dot{\gamma_{i}}}} \right) &=& \displaystyle\lim_{\dot{\gamma_{i}} \rightarrow 0} \ln \left( \dfrac{g_{\dot{\gamma_{i}}}}{f_{\dot{\gamma_{i}}}(1 + g_{\dot{\gamma_{i}}})} \right)^{g_{\dot{\gamma_{i}}}} \\
        \\
        &=& \displaystyle\lim_{\dot{\gamma_{i}} \rightarrow 0} g_{\dot{\gamma_{i}}} \cdot \ln \left( \dfrac{g_{\dot{\gamma_{i}}}}{f_{\dot{\gamma_{i}}}} \right) \\
        \\
        &=& \displaystyle\lim_{\dot{\gamma_{i}} \rightarrow 0} g_{\dot{\gamma_{i}}} \cdot \left( \ln g_{\dot{\gamma_{i}}} - \ln f_{\dot{\gamma_{i}}} \right) ; \\
        \\
        \text{therefore, } |\ln g_{\dot{\gamma_{i}}} - \ln f_{\dot{\gamma_{i}}}| < \infty & \rightarrow & \displaystyle\lim_{\dot{\gamma_{i}} \rightarrow 0} g_{\dot{\gamma_{i}}} \cdot \left( \ln g_{\dot{\gamma_{i}}} - \ln f_{\dot{\gamma_{i}}} \right) = 0 \\
        \\
        & \rightarrow & \displaystyle\lim_{\dot{\gamma_{i}} \rightarrow 0} \left( \dfrac{g_{\dot{\gamma_{i}}}}{f_{\dot{\gamma_{i}}}(1 + g_{\dot{\gamma_{i}}})} \right)^{g_{\dot{\gamma_{i}}}} = 1 . \\
      \end{array}
    \end{equation*}

    Since we also have the condition $\lim_{\dot{\gamma_{i}} \rightarrow 0} \left( \frac{|\tau_{2} - \tau_{1}|}{(1 + z_{i})\tau_{i}} \right)^{g(\dot{\gamma_{i}})} = 1$ with respect to the other country-level variables, we finally conclude that $\lim_{\dot{\gamma_{i}} \rightarrow 0} p^{*}(\dot{\gamma_{i}}, \cdot) = \bar{p}^{c} + 1(p^{c} - \bar{p}^{c}) = p^{c}$.
  \end{proof}
\end{proposition}

\emph{Proposition \ref{proposition lim optimal p in enforcement}} confirms that the maximising transfer prices at the boundaries of the tax enforcement domain, $\dot{\gamma_{i}} \rightarrow 0$ and $\dot{\gamma_{i}} \rightarrow 1$ are both $\alpha$-cuts of the fuzzy arm's length price $\tilde{P}^{c}$. Nonetheless, for marginal changes within the domain interval $\dot{\gamma_{i}} \in (0,1]$, the effect over the optimal transfer price $p^{*}(\dot{\gamma_{i}}, \cdot)$ depends on how functions $g_{\dot{\gamma_{i}}}$, $f_{\dot{\gamma_{i}}}$ vary. Differentiating $p^{*}(\dot{\gamma_{i}}, \cdot)$ with respect to $\dot{\gamma_{i}}$, we obtain the following standard-form equations:

\begin{equation} \label{dp d dot gamma}
  \dfrac{\partial p^{*}(\dot{\gamma_{i}}, \cdot)}{\partial \dot{\gamma_{i}}} = \Delta p^{*} \cdot \left[ \dfrac{d g_{\dot{\gamma_{i}}}}{d \dot{\gamma_{i}}} \cdot \left( \ln \left( \dfrac{|\tau_{2} - \tau_{1}|}{f_{\dot{\gamma_{i}}} \cdot (1 + z_{i}) \cdot \tau_{i} \cdot \left( 1 + \frac{1}{g_{\dot{\gamma_{i}}}} \right)} \right) + \dfrac{1}{1 + g_{\dot{\gamma_{i}}}} \right) - \dfrac{d f_{\dot{\gamma_{i}}}}{d \dot{\gamma_{i}}} \cdot \dfrac{g_{\dot{\gamma_{i}}}}{f_{\dot{\gamma_{i}}}} \right] ;
\end{equation}

For Eq. \ref{dp d dot gamma}, we observe again that the variation $\partial p^{*}(\dot{\gamma_{i}}, \cdot) / \partial \dot{\gamma_{i}}$ is not necessarily negative for all $\forall \dot{\gamma_{i}} \in (0,1]$ -- compare it with Eq. \ref{optimal dp dgamma}. A negative variation such that $sgn(\partial p^{*}(\dot{\gamma_{i}}, \cdot) / \partial \dot{\gamma_{i}}) = -sgn(\Delta p^{*}(\dot{\gamma_{i}}, \cdot))$ requires the following necessary condition for the functions $g_{\dot{\gamma_{i}}}$, $f_{\dot{\gamma_{i}}}$:

\begin{proposition} \label{prop f condition}
  Variation in the optimal transfer price $p^{*}(\dot{\gamma_{i}}, \cdot)$ with respect to marginal changes in the tax enforcement $\dot{\gamma_{i}}$ is opposite to the profit shifting incentive $\tau_{2} - \tau_{1}$, such that $sgn(\partial p^{*}(\dot{\gamma_{i}}, \cdot) / \partial \dot{\gamma_{i}}) = -sgn(\Delta p^{*}(\dot{\gamma_{i}}, \cdot))$, iff the functions $g_{\dot{\gamma_{i}}}$, $f_{\dot{\gamma_{i}}}$ satisfy the condition
  \begin{equation*}
    \forall \dot{\gamma_{i}} \in (0,1] : f_{\dot{\gamma_{i}}} \geq \dfrac{g_{\dot{\gamma_{i}}}}{1 + g_{\dot{\gamma_{i}}}} .
  \end{equation*}
  
  \begin{proof}
    
    For simplification, assume initially the equalities $\tau_{j \neq i} = 0$, $z_{i} = 0$. For the Eq. \ref{dp d dot gamma} to have a negative effect, such that $sgn(\partial p^{*}(\dot{\gamma_{i}}, \cdot) / \partial \dot{\gamma_{i}}) = -sgn(\Delta p^{*}(\dot{\gamma_{i}}, \cdot))$, it requires the necessary condition
    
    \begin{equation*}
      \dfrac{d g_{\dot{\gamma_{i}}}}{d \dot{\gamma_{i}}} \cdot \left( \ln \left( \dfrac{1}{f_{\dot{\gamma_{i}}} \cdot \left( 1 + \frac{1}{g_{\dot{\gamma_{i}}}} \right)} \right) + \dfrac{1}{1 + g_{\dot{\gamma_{i}}}} \right) - \dfrac{d f_{\dot{\gamma_{i}}}}{d \dot{\gamma_{i}}} \cdot \dfrac{g_{\dot{\gamma_{i}}}}{f_{\dot{\gamma_{i}}}} \leq 0
    \end{equation*}
    
    \noindent for the complete domain $\forall \dot{\gamma_{i}} \in (0,1]$. Rearranging, we obtain the inequality
    
    \begin{equation*}
      \dfrac{d f_{\dot{\gamma_{i}}}}{d g_{\dot{\gamma_{i}}}} \cdot \dfrac{g_{\dot{\gamma_{i}}}}{f_{\dot{\gamma_{i}}}} \geq \ln \left( \dfrac{g_{\dot{\gamma_{i}}}}{f_{\dot{\gamma_{i}}} \cdot ( 1 + g_{\dot{\gamma_{i}}} )} \right) + \dfrac{1}{1 + g_{\dot{\gamma_{i}}}} .
    \end{equation*}
    
    At the critical point $\partial p^{*}(\dot{\gamma_{i}}, \cdot) / \partial \dot{\gamma_{i}} = 0$, we clearly have
    
    \begin{equation*}
      f_{\dot{\gamma_{i}}} = \dfrac{g_{\dot{\gamma_{i}}}}{1 + g_{\dot{\gamma_{i}}}} . \\
    \end{equation*}
    
    Now, for any small perturbation $\delta \neq 0$ on function $f_{\dot{\gamma_{i}}}$ such that we have $\delta \neq 0 : f_{\dot{\gamma_{i}}} = g_{\dot{\gamma_{i}}}/(1 + g_{\dot{\gamma_{i}}}) + \delta$, the necessary condition is equal to
    \begin{equation*}
      \begin{array}{rcl}
        \dfrac{d}{d g_{\dot{\gamma_{i}}}} \left( \dfrac{g_{\dot{\gamma_{i}}}}{1 + g_{\dot{\gamma_{i}}}} + \delta \right) \cdot \left( \dfrac{g_{\dot{\gamma_{i}}}(1 + g_{\dot{\gamma_{i}}})}{g_{\dot{\gamma_{i}}} + \delta (1 + g_{\dot{\gamma_{i}}})} \right) & \geq & \ln \left( \dfrac{g_{\dot{\gamma_{i}}}}{g_{\dot{\gamma_{i}}} + \delta(1 + g_{\dot{\gamma_{i}}})} \right) + \dfrac{1}{1 + g_{\dot{\gamma_{i}}}} \\
        \\
        \dfrac{1}{1 + g_{\dot{\gamma_{i}}}} \cdot \dfrac{g_{\dot{\gamma_{i}}}}{g_{\dot{\gamma_{i}}} + \delta(1 + g_{\dot{\gamma_{i}}})} & \geq & \ln \left( \dfrac{g_{\dot{\gamma_{i}}}}{g_{\dot{\gamma_{i}}} + \delta(1 + g_{\dot{\gamma_{i}}})} \right) + \dfrac{1}{1 + g_{\dot{\gamma_{i}}}} \\
        \\
        \dfrac{1}{1 + g_{\dot{\gamma_{i}}}} \cdot \left( \dfrac{g_{\dot{\gamma_{i}}} - (g_{\dot{\gamma_{i}}} + \delta(1 + g_{\dot{\gamma_{i}}}))}{g_{\dot{\gamma_{i}}} + \delta(1 + g_{\dot{\gamma_{i}}})} \right) & \geq & \ln \left( \dfrac{g_{\dot{\gamma_{i}}}}{g_{\dot{\gamma_{i}}} + \delta(1 + g_{\dot{\gamma_{i}}})} \right) \\
        \\
        \dfrac{- \delta}{g_{\dot{\gamma_{i}}} + \delta(1 + g_{\dot{\gamma_{i}}})} & \geq & \ln \left( \dfrac{g_{\dot{\gamma_{i}}}}{g_{\dot{\gamma_{i}}} + \delta(1 + g_{\dot{\gamma_{i}}})} \right) . \\
      \end{array}
    \end{equation*}
    
    The inequality is satisfied for any non-negative value $\delta \geq 0$. On the other hand, if the perturbation is negative such that $\delta < 0 \rightarrow f_{\dot{\gamma_{i}}} < g_{\dot{\gamma_{i}}}/(1 + g_{\dot{\gamma_{i}}})$, we arrive at a contradiction -- the right hand side of the inequality becomes the larger term. Hence, it implies
    
    \begin{equation*}
      \left \{ \delta \geq 0 : f_{\dot{\gamma_{i}}} = \dfrac{g_{\dot{\gamma_{i}}}}{1 + g_{\dot{\gamma_{i}}}} + \delta \right \} \rightarrow f_{\dot{\gamma_{i}}} \geq \dfrac{g_{\dot{\gamma_{i}}}}{1 + g_{\dot{\gamma_{i}}}} .
    \end{equation*}
    
    At last, this necessary condition is clearly maintained if we drop the simplifications $\tau_{j \neq i} = 0$, $z_{i} = 0$; for both LTP and HTP cases, $\tau_{1} \neq \tau_{2}$, $\tau_{j \neq i} > 0$, $z_{i} > 0$, we have the inequality
    
    \begin{equation*}
      \dfrac{- \delta}{g_{\dot{\gamma_{i}}} + \delta(1 + g_{\dot{\gamma_{i}}})} \geq \ln \left( \dfrac{g_{\dot{\gamma_{i}}}}{g_{\dot{\gamma_{i}}} + \delta(1 + g_{\dot{\gamma_{i}}})} \right) > \ln \left( \dfrac{g_{\dot{\gamma_{i}}}}{g_{\dot{\gamma_{i}}} + \delta(1 + g_{\dot{\gamma_{i}}})} \right) + \ln \left( \dfrac{|\tau_{2} - \tau_{1}|}{(1 + z_{i}) \cdot \tau_{i}} \right)
    \end{equation*}
    
    \noindent which also implies $f_{\dot{\gamma_{i}}} \geq g_{\dot{\gamma_{i}}}/(1 + g_{\dot{\gamma_{i}}})$.
  \end{proof}
\end{proposition}

\emph{Proposition \ref{proposition lim optimal p in enforcement}} shows that the optimal transfer price at the lowest enforcement level, $\dot{\gamma_{i}} \rightarrow 0$ is equal to the least tolerable arm's length price $p^{c}$, e.g. for the lowest tax enforcement, the MNE may choose the transfer price equal to $p^{c} \in \tilde{P}^{c}$, which is the farthest from the mode $\bar{p}^{c}$. Moreover, \emph{Proposition \ref{prop f condition}} presents the necessary condition for the variation $\partial p^{*}(\dot{\gamma_{i}}, \cdot) / \partial \dot{\gamma_{i}}$ to produce an effect opposite to the profit shifting incentive, $\tau_{2} - \tau_{1}$ through the complete domain $\dot{\gamma_{i}} \in (0,1]$. Hence, it implies the following:

\begin{corollary} \label{cor f condition}
  \emph{Proposition \ref{prop lambda condition}} and \emph{Proposition \ref{prop f condition}} are equivalent.
  \begin{proof}
    Derives directly from Eq. \ref{nonhom audit intensity} and \emph{Proposition \ref{prop lambda condition}}, with the simplifications $\tau_{j \neq i} = 0$, $z_{i} = 0$, for we have 
    \begin{equation*}
      \begin{array}{rcl}
        \Lambda_{i}(0, \dot{\gamma_{i}}) = -\ln(1 - f(\dot{\gamma_{i}})) &\geq& -\ln \left( 1 - \dfrac{1}{\left(1 + \frac{1}{g(\dot{\gamma_{i}})} \right)} \right) \\
        \\
        1 - f(\dot{\gamma_{i}}) & \leq & \dfrac{1}{1 + g(\dot{\gamma_{i}})} \\
        \\
        f(\dot{\gamma_{i}}) & \geq & \dfrac{g(\dot{\gamma_{i}})}{1 + g(\dot{\gamma_{i}})} . \\
      \end{array}
    \end{equation*}
  \end{proof}
\end{corollary}

From \emph{Corollary \ref{cor f condition}}, it means that the inequality $f_{\dot{\gamma_{i}}} \geq g_{\dot{\gamma_{i}}} / (1 + g_{\dot{\gamma_{i}}})$ is also a necessary condition for the optimal transfer price $p^{*}(\dot{\gamma_{i}}, \cdot)$ to be a $\alpha$-cut of the fuzzy arm's length price $\tilde{P}^{c}$. Otherwise, we may have a tax enforcement level such that $\exists \dot{\gamma_{i}} \in (0,1] : p^{*}(\dot{\gamma_{i}}, \cdot) \notin \tilde{P}^{c}$.

Now, we derive a sufficient condition for the optimal transfer price $p^{*}(\dot{\gamma_{i}}, \cdot)$ to be a $\alpha$-cut of the fuzzy arm's length price $\tilde{P}^{c}$ for the complete domain $\forall \dot{\gamma_{i}} \in (0,1]$.

\begin{proposition} \label{prop f sufficient condition}
  If the functions $g_{\dot{\gamma_{i}}}$, $f_{\dot{\gamma_{i}}}$ satisfy the condition $f_{\dot{\gamma_{i}}} \geq g_{\dot{\gamma_{i}}}$ for the complete domain $\forall \dot{\gamma_{i}} \in (0,1]$, the optimal transfer price $p^{*}(\dot{\gamma_{i}}, \cdot)$ is a $\alpha$-cut of the fuzzy arm's length price $\tilde{P}^{c}$.
  \begin{proof}
    \emph{Corollary \ref{cor f condition}} derives the necessary condition $f_{\dot{\gamma_{i}}} \geq g_{\dot{\gamma_{i}}}/(1 + g_{\dot{\gamma_{i}}})$ for the optimal transfer price $p^{*}(\dot{\gamma_{i}}, \cdot)$ to be a $\alpha$-cut of the fuzzy arm's length price $\tilde{P}^{c}$ with respect to the complete domain $\forall \dot{\gamma_{i}} \in (0,1]$. First, we are sure that the necessary condition attains the equality at the lower bound of the domain, $\dot{\gamma_{i}} \rightarrow 0$, for we have
    
    \begin{equation*}
      \begin{array}{rcl}
        \displaystyle\lim_{\dot{\gamma_{i}} \to 0} \left( f_{\dot{\gamma_{i}}} \geq \dfrac{g_{\dot{\gamma_{i}}}}{1 + g_{\dot{\gamma_{i}}}} \right) &=& \displaystyle\lim_{\dot{\gamma_{i}} \to 0} f_{\dot{\gamma_{i}}} \geq \displaystyle\lim_{\dot{\gamma_{i}} \to 0} \left( \dfrac{g_{\dot{\gamma_{i}}}}{1 + g_{\dot{\gamma_{i}}}} \right) \\
        \\
        &=& \displaystyle\lim_{\dot{\gamma_{i}} \to 0} f_{\dot{\gamma_{i}}} = \displaystyle\lim_{\dot{\gamma_{i}} \to 0} \left( \dfrac{g_{\dot{\gamma_{i}}}}{1 + g_{\dot{\gamma_{i}}}} \right) = 0 . \\
      \end{array}
    \end{equation*}
    
    But we also have that $\lim_{\dot{\gamma_{i}} \to 0} g_{\dot{\gamma_{i}}}/(1 + g_{\dot{\gamma_{i}}}) = \lim_{\dot{\gamma_{i}} \to 0} g_{\dot{\gamma_{i}}} = 0$, so all terms converge to zero at the very initial point $\dot{\gamma_{i}} \rightarrow 0$:
    
    \begin{equation*}
      \lim_{\dot{\gamma_{i}} \to 0} g_{\dot{\gamma_{i}}} = \lim_{\dot{\gamma_{i}} \to 0} f_{\dot{\gamma_{i}}} = \lim_{\dot{\gamma_{i}} \to 0}  \left( \dfrac{g_{\dot{\gamma_{i}}}}{1 + g_{\dot{\gamma_{i}}}} \right) = 0 .
    \end{equation*}
    
    However, as the tax enforcement increases, $\dot{\gamma_{i}} \in (0,1] : \dot{\gamma_{i}} > 0$, this equality is not maintained, since it implies $\dot{\gamma_{i}} > 0 \rightarrow g_{\dot{\gamma_{i}}} > g_{\dot{\gamma_{i}}}/(1 + g_{\dot{\gamma_{i}}})$. Besides, as the tax enforcement reaches the upper bound of the domain, $\dot{\gamma_{i}} \rightarrow 1$, we have $\lim_{\dot{\gamma_{i}} \to 1} g_{\dot{\gamma_{i}}} = \lim_{\dot{\gamma_{i}} \to 1} f_{\dot{\gamma_{i}}} = 1$, thus it implies $\lim_{\dot{\gamma_{i}} \to 1} f_{\dot{\gamma_{i}}} > \lim_{\dot{\gamma_{i}} \to 1} ( g_{\dot{\gamma_{i}}}/(1 + g_{\dot{\gamma_{i}}}))$. Combining these two cases, we derive two possible inequalities:
    
    \begin{equation*}
      \left \{ \dot{\gamma_{i}} \in (0,1] : \dot{\gamma_{i}} > 0 \right \} \rightarrow \left \{
      \begin{array}{l}
        g_{\dot{\gamma_{i}}} \geq f_{\dot{\gamma_{i}}} \geq \dfrac{g_{\dot{\gamma_{i}}}}{1 + g_{\dot{\gamma_{i}}}} ; \\
        \\
        f_{\dot{\gamma_{i}}} \geq g_{\dot{\gamma_{i}}} > \dfrac{g_{\dot{\gamma_{i}}}}{1 + g_{\dot{\gamma_{i}}}} . \\
      \end{array} \right.
    \end{equation*}
    
    It shows that if we have $g_{\dot{\gamma_{i}}} \geq f_{\dot{\gamma_{i}}}$, we still need to confirm that the necessary condition $f_{\dot{\gamma_{i}}} \geq g_{\dot{\gamma_{i}}}/(1 + g_{\dot{\gamma_{i}}})$ is satisfied. On the other hand, if we have $f_{\dot{\gamma_{i}}} \geq g_{\dot{\gamma_{i}}}$, the necessary condition in \emph{Corollary \ref{cor f condition}} is automatically satisfied. Therefore, the condition $\forall \dot{\gamma_{i}} \in (0,1] : f_{\dot{\gamma_{i}}} \geq g_{\dot{\gamma_{i}}}$ is a sufficient condition for the optimal transfer price $p^{*}(\dot{\gamma_{i}}, \cdot)$ is a $\alpha$-cut of the fuzzy arm's length price $\tilde{P}^{c}$.
       
  \end{proof}
\end{proposition}


Remark that functions $g_{\dot{\gamma_{i}}}$, $f_{\dot{\gamma_{i}}}$ refer to the arm's length tolerance parameter and the probability of tax audit respectively. \emph{Proposition \ref{prop f sufficient condition}} thus indicates that the optimal transfer price $p^{*}(\dot{\gamma_{i}}, \cdot)$ is certainly a $\alpha$-cut of the fuzzy arm's length price $\tilde{P}^{c}$ if the audit probability is greater than the arm's length tolerance parameter, for the complete domain $\forall \dot{\gamma_{i}} \in (0,1]$. Otherwise, we still need to confirm that the necessary condition $f_{\dot{\gamma_{i}}} \geq g_{\dot{\gamma_{i}}}/(1 + g_{\dot{\gamma_{i}}})$ is satisfied.

\section{Discussion and Conclusion} \label{conclusion}

This paper presents a model for optimal tax-induced transfer pricing under fuzzy arm's length parameter. The fuzzy arm's length price follows the structure of a fuzzy number \cite{zadeh1965} by means of a concave shape function with smooth membership grading, which varies with respect to the arm's length tolerance parameter of tax authorities. Under usual conditions, the optimal transfer price becomes a maximising $\alpha$-cut of the fuzzy arm's length price, while it still satisfies the conventional assumptions of convex concealment costs and increasing profit-shifting incentives at an increasing tax differentials.

At first, we show that the MNE always obtains a gain from profit shifting up to the optimal transfer price, regardless of the shifting direction, and this gain is obtained at any levels of tax penalty, audit probability and arm's length tolerance. Gains from profit shifting are intensified by adjusting the intra-firm outputs at a second maximisation stage. Moreover, the MNE may obtain exceeding gains by extrapolating the fuzzy arm's length parameter if the probability of tax audits is sufficiently low. This extreme case is prevented specially by increasing the audit intensity or intensifying the other anti-shifting mechanisms on the harmed country.

These analyses offer some interesting insights on how the ambiguity of the arm's length parameter may affect the profit shifting strategy of firms. First and foremost, the fuzziness of the arm's length parameter can be used by firms to achieve their profit shifting goals, since this fuzziness is the condition that implies the gains from profit shifting. For any questioning by the tax authority, the transfer price may be more or less sustained based on arguments about the conditions of the comparable transactions. Moreover, even if the tax authority observes all intra-firm transactions in a full-audit mode, the ambiguity of what can considered an appropriate transfer price is not eliminated. It means that the uncertainty is not attributed only to the probability of being audited, but also on the tolerance level of the tax auditor. And this uncertainty can be beneficial for firms focusing on a profit shifting strategy. At last, anti-shifting rules impose the arm's length criterion for the transfer prices, but no requirements are currently imposed for the level of internal outputs. In this sense, any change in the arm's length tolerance of tax authorities might be offset by adjustments in internal outputs if the MNE has some operational flexibility, so the final amount of shifted profits remains the same.

\section*{Appendix} \label{Appendix}

\subsection*{Derivation of Eq. \ref{prob tax audit}} \label{derivation gamma}

For a variable $ g \in \mathbb{R}_{++}$ as any point within a continuum of occurrences at a constant average rate $\lambda$, the time $y$ of the $g$-th occurrence is a random variable that follows a gamma distribution and has a cumulative probability as $ \mathbb{P}(y) = 1 - \Gamma (g, \lambda) / \Gamma(g)$. The gamma function is defined as
 
\begin{equation*}
 \Gamma (g) = \int_{0}^{\infty} \lambda^{g} y^{g-1} e^{-\lambda y} \, \mathrm{d}y
\end{equation*}

\noindent and the upper gamma function is defined as

\begin{equation*}
 \Gamma (g, \lambda) = \int_{\lambda}^{\infty} \lambda^{g} y^{g-1} e^{-\lambda y} \, \mathrm{d}y .
\end{equation*}

Since we have discrete events as $g=k : k \in \mathbb{N}$, the upper gamma function can be expressed as the series expansion

\begin{equation*}
 \Gamma (k, \lambda) = (k-1)! \cdot e^{-\lambda y} \sum_{q = 0}^{k-1} \frac{(\lambda y)^{q}}{q!} .
\end{equation*}

The second multiplier at the right hand side of the above equation represents the cumulative chance of $k-1$ events to occur at intensity $\lambda$ up to moment $y$, where $k$ is a Poisson random variable\footnote{The relation $\mathbb{P}(y)=1-\mathbb{P}(k-1)$ indicates that changes in occurrence rate $\lambda$ produce an inverse impact on the distribution; a random gamma-distributed variable $y$ has mean $\mathbb{E}(y) = k/\lambda$ and variance $\mathbb{V}(y) = k/\lambda^{2}$.}. In addition, the gamma function satisfy the property $\Gamma(k)= (k-1)!$ for discrete variables $k \in \mathbb{N}$, which implies the equality $k\Gamma(k) = \Gamma(k+1)$. Hence, assuming the occurrence rate $\lambda$ is obtained for a period up to $y$, thus $y=1$, these conditions allow us to derive the cumulative probability distribution of $k$ events as

\begin{equation*}
 \sum_{q = 0}^k \mathbb{P}(q) = \frac{k!}{k!} \cdot \sum_{q = 0}^k \frac{\lambda^{q} e^{-\lambda}}{q!} = \frac{\Gamma (k+1, \lambda)}{\Gamma(k+1)}
\end{equation*}

\noindent which is presented in Equation \ref{prob tax audit}. Poisson cumulative distribution function $\sum_{q = 0}^k \mathbb{P}(q)$ expressed by means of gamma function $\Gamma (k,\lambda)$ is defined for all positive real numbers and provides continuity condition for the analysis.

\subsection*{Derivation of Eq. \ref{optimal dp dt}-\ref{optimal ddp dtt}} \label{derivation dp dt}

For all real numbers $\forall x \in \mathbb{R}$, $x$ is equal to $x = sgn(x) \cdot |x|$, with $|\cdot| : \mathbb{R} \rightarrow \mathbb{R}_{+}$ as the absolute value function and $sgn(x)$ as the sign function satisfying

\begin{equation*}
 sgn(x) = \left\{
  \begin{array}{rl}
  -1, & \text{iff } x<0\\
  0, & \text{iff } x=0\\
  1, & \text{iff } x>0 .\\
  \end{array} \right.
\end{equation*}

For $\forall x \neq 0$, we have $sgn(x) = x / |x| = |x| / x \rightarrow |x| = x \cdot sgn(x) = x / sgn(x)$, which implies $\partial |x| / \partial x = sgn(x)$. Differentiating $|\tau_{2} - \tau_{1}|$ with respect to $\tau_{i}$ for both LTP and HTP cases provides

\begin{equation*}
  \dfrac{\partial |\tau_{2} - \tau_{1}|}{\partial \tau_{i}} = \left\{
   \begin{array}{rclr}
    \text{LTP } \rightarrow \{ \tau_{2} < \tau_{1}, i=1 \} & \rightarrow & \frac{\partial |\tau_{2} - \tau_{1}|}{\partial \tau_{1}} =& sgn(\tau_{2} - \tau_{1}) \cdot \frac{\partial (\tau_{2} - \tau_{1})}{\partial \tau_{1}} = 1 ; \\
    \\
    \text{HTP } \rightarrow \{ \tau_{2} > \tau_{1}, i=2 \} & \rightarrow & \frac{\partial |\tau_{2} - \tau_{1}|}{\partial \tau_{2}} =& sgn(\tau_{2} - \tau_{1}) \cdot \frac{\partial (\tau_{2} - \tau_{1})}{\partial \tau_{2}} = 1 .
   \end{array} \right.
\end{equation*}

Under these properties, differentiating Eq. \ref{optimal p} with respect to $\tau_{i}$ provides the following standard form:

\begin{equation*}
 \begin{array}{rcl}
  \dfrac{\partial p^{*}}{\partial \tau_{i}} &=& \dfrac{\partial \bar{p}^{c}}{\partial \tau_{i}} + \dfrac{\partial \Delta p^{*}}{\partial \tau_{i}} \\
  \\
  &=& 0 + \gamma_{i} \left( \dfrac{|\tau_{2} - \tau_{1}|}{(1 - e^{-\lambda_{i}}) (1 + z_{i}) \left(1 + \frac{1}{\gamma_{i}} \right) \cdot \tau_{i}} \right)^{\gamma_{i} - 1} \cdot \dfrac{\partial}{\partial \tau_{i}} \left( \dfrac{|\tau_{2} - \tau_{1}|}{(1 - e^{-\lambda_{i}}) (1 + z_{i}) \left(1 + \frac{1}{\gamma_{i}} \right) \cdot \tau_{i}} \right) (p^{c} - \bar{p}^{c}) \\
  \\
  &=& \gamma_{i} \cdot \dfrac{|\tau_{2} - \tau_{1}|^{\gamma_{i} - 1}}{\left( (1 - e^{-\lambda_{i}}) (1 + z_{i}) \left(1 + \frac{1}{\gamma_{i}} \right) \cdot \tau_{i} \right)^{\gamma_{1} + 1}} \cdot \left[ \tau_{i} \cdot (1 - e^{-\lambda_{i}}) (1 + z_{i}) \left(1 + \frac{1}{\gamma_{i}} \right) \right.\\
  \\
  && \left.- |\tau_{2} - \tau_{1}| \cdot (1 - e^{-\lambda_{i}}) (1 + z_{i}) \left(1 + \frac{1}{\gamma_{i}} \right) \right](p^{c} - \bar{p}^{c}) \\
  \\
  &=& \dfrac{\gamma_{i}(p^{c} - \bar{p}^{c})}{\tau_{i}^{\gamma_{i} + 1} \cdot |\tau_{2} - \tau_{1}|^{1 - \gamma_{i}}} \cdot \dfrac{ (1 - e^{-\lambda_{i}}) (1 + z_{i}) \left(1 + \frac{1}{\gamma_{i}} \right)}{\left( (1 - e^{-\lambda_{i}}) (1 + z_{i}) \left(1 + \frac{1}{\gamma_{i}} \right) \right)^{\gamma_{i} + 1}} \cdot (\tau_{i} - |\tau_{2} - \tau_{1}|) \\
  \\
  &=& \dfrac{\gamma_{i} (p^{c} - \bar{p}^{c})}{\left( (1 - e^{-\lambda_{i}}) \cdot (1 + z_{i}) \cdot \left(1 + \frac{1}{\gamma_{i}} \right) \right)^{\gamma_{i}}} \cdot \dfrac{\tau_{i} - |\tau_{2} - \tau_{1}|}{\tau_{i}^{1 + \gamma_{i}} \cdot |\tau_{2} - \tau_{1}|^{1 - \gamma_{i}}} . \\
  \end{array}
\end{equation*}

For $\partial^{2} p^{*} / \partial \tau_{i}^{2}$, simplify the constant multiplier $\Xi_{i}^{c} = \frac{\gamma_{i} (p^{c} - \bar{p}^{c})}{\left( (1 - e^{-\lambda_{i}}) \cdot (1 + z_{i}) \cdot \left(1 + \frac{1}{\gamma_{i}} \right) \right)^{\gamma_{i}}}$ at the right hand side of Eq. \ref{optimal dp dt}. Standard form in Eq. \ref{optimal ddp dtt} is derived as follows:
  
\begin{equation*}
 \begin{array}{rcl}
  \dfrac{\partial^{2} p^{*}}{\partial \tau_{i}^{2}} &=& \dfrac{\gamma_{i} (p^{c} - \bar{p}^{c})}{\left( (1 - e^{\lambda_{i}}) \cdot (1 + z_{i}) \cdot \left( 1 + \frac{1}{\gamma_{i}} \right) \right)^{\gamma_{i}}} \cdot \dfrac{d}{d \tau_{i}} \left( \dfrac{\tau_{i} - |\tau_{2} - \tau_{1}|}{\tau_{i}^{1 + \gamma_{i}} \cdot |\tau_{2} - \tau_{1}|^{1 - \gamma_{i}}} \right) \\
  \\
  &=& \Xi_{i}^{c} \cdot \left( 0 - \dfrac{(\tau_{i} - |\tau_{2} - \tau_{1}|) \cdot [d(\tau_{i}^{1 + \gamma_{i}} \cdot |\tau_{2} - \tau_{1}|^{1 - \gamma_{i}})/ d \tau_{i}]}{(\tau_{i}^{1 + \gamma_{i}} \cdot |\tau_{2} - \tau_{1}|^{1 - \gamma_{i}})^{2}} \right) \\
  \\
  &=& \Xi_{i}^{c} \cdot \left[ \left( - \dfrac{\tau_{i} - |\tau_{2} - \tau_{1}|}{\tau_{i}^{1 + \gamma_{i}} \cdot |\tau_{2} - \tau_{1}|^{1 - \gamma_{i}}} \right) \cdot \left[ \left( \dfrac{(1 + \gamma_{i})\tau_{i}^{\gamma_{i}} \cdot |\tau_{2} - \tau_{1}|^{1 - \gamma_{i}}}{\tau_{i}^{1 + \gamma_{i}} \cdot |\tau_{2} - \tau_{1}|^{1 - \gamma_{i}}} \right) \right. \right. \\
  \\
  && \left. \left. + \left( \dfrac{(1 - \gamma_{i}) \cdot |\tau_{2} - \tau_{1}|^{-\gamma_{i}} \cdot \tau_{i}^{1 + \gamma_{i}}}{\tau_{i}^{1 + \gamma_{i}} \cdot |\tau_{2} - \tau_{1}|^{1 - \gamma_{i}}} \right) \right] \right] \\
  \\
  &=& \Xi_{i}^{c} \cdot \left[ - \left( \dfrac{\tau_{i} - |\tau_{2} - \tau_{1}|}{\tau_{i}^{1 + \gamma_{i}} \cdot |\tau_{2} - \tau_{1}|^{1 - \gamma_{i}}} \right) \cdot \left( \dfrac{1 + \gamma_{i}}{\tau_{i}} + \dfrac{1 - \gamma_{i}}{|\tau_{2} - \tau_{1}|} \right) \right] \\
  \\
  &=& \Xi_{i}^{c} \cdot \left( \dfrac{\tau_{i} - |\tau_{2} - \tau_{1}|}{\tau_{i}^{1 + \gamma_{i}} \cdot |\tau_{2} - \tau_{1}|^{1 - \gamma_{i}}} \right) \cdot \left( \dfrac{(\gamma_{i} - 1) \cdot \tau_{i} - (\gamma_{i} + 1) \cdot |\tau_{2} - \tau_{1}|}{\tau_{i} \cdot |\tau_{2} - \tau_{1}|} \right) \\
  \\
  &=& \dfrac{\partial p^{*}}{\partial \tau_{i}} \cdot \dfrac{(\gamma_{i} - 1) \cdot \tau_{i} - (\gamma_{i} + 1) \cdot |\tau_{2} - \tau_{1}|}{\tau_{i} \cdot |\tau_{2} - \tau_{1}|} \\
 \end{array}
\end{equation*}

From the $\tau_{i}$-semi-elasticity of the optimal price gap $\Delta p^{*}$ equal to $\varepsilon_{\Delta p^{*}, \tau_{i}} / \tau_{i} = \gamma_{i} \cdot \frac{\tau_{i} - |\tau_{2} - \tau_{1}|}{\tau_{i} \cdot |\tau_{2} - \tau_{1}|}$, the second-order $\tau_{i}$-semi-elasticity of $p^{*}$ is defined as

\begin{equation*}
 \begin{array}{rcl}
  \dfrac{\varepsilon^{2}_{p^{*}, \tau_{i}}}{\tau_{i}} &=& \dfrac{\partial^{2} p^{*} / \partial \tau_{i}^{2}}{\partial p^{*} / \partial \tau_{i}} \\
  \\
  &=& \dfrac{\varepsilon_{\Delta p^{*}, \tau_{i}}}{\tau_{i}} - \dfrac{\tau_{i} - |\tau_{2} - \tau_{1}|}{\tau_{i} \cdot |\tau_{2} - \tau_{1}|} \\
  \\
  &=& \dfrac{(\gamma_{i} - 1) \cdot \tau_{i} - (\gamma_{i} + 1) \cdot |\tau_{2} - \tau_{1}|}{\tau_{i} \cdot |\tau_{2} - \tau_{1}|}
 \end{array}
\end{equation*}

\noindent which is the multiplier at the right hand side of Eq. \ref{optimal ddp dtt}.

\subsection*{Derivation of Eq. \ref{prob tax audit parametric in enforcement}} \label{derivation prob in tax enforcement}

The homogeneous audit intensity $\lambda_{i} > 0$ can be any positive real number, thus the range of the corresponding non-homogeneous audit rate function $\lambda_{i}(\dot{\gamma_{i}}) : (0,1] \rightarrow \mathbb{R}_{++}$ is unbounded above. Since we have a bounded domain $\dot{\gamma_{i}} \in (0,1]$, therefore the rate function $\lambda_{i}(\dot{\gamma_{i}})$ must indeed be unbounded above\footnote{Of course, the condition for the function $\lambda_{i}(\dot{\gamma_{i}})$ to be unbounded above necessarily arises from our restriction of the domain of $\dot{\gamma_{i}}$ to the bounded interval $(0,1]$.}, i.e. under the definition of non-uniformly continuous functions, it implies

\begin{equation*}
  \{ \forall \delta > 0 \text{, } \dot{\gamma_{i}} \in (0,1] : |\Delta \dot{\gamma_{i}}| < \delta \} \rightarrow \lim_{\dot{\gamma_{i}} \rightarrow 1} | \lambda_{i}(\dot{\gamma_{i}} + \Delta \dot{\gamma_{i}}) - \lambda_{i}(\dot{\gamma_{i}})| = \infty.
\end{equation*}

From the general function $f(\dot{\gamma_{i}}) : (0,1] \rightarrow \mathbb{R}_{++}$ as defined in Section \ref{Modelling a General Tax Enforcement Effect}, let the audit rate function follow a negative semi-elasticity design\footnote{For a differentiable function $f(y) : Y \rightarrow \mathbb{R}$, $y \in Y$, $Y$ is bounded, the negative semi-elasticity equal to $- \frac{d f(y)}{d y} \cdot \frac{1}{f(y)}$ is unbounded above at the zeros $f(y) \rightarrow 0$. Two classical examples are the functions $-1/y$ and $\tan(y)$, which are unbounded above on the bounded domains $y \in [-1,0]$ and $y \in [0,\pi/2]$ respectively. Both examples are defined as the negative semi-elasticities of the functions $f(y) = -y$ and $f(y) = \cos(y)$ as follows:

\begin{equation*}
  f(y) = -y : - \dfrac{d (-y)}{d y} \cdot \dfrac{1}{-y} = - \dfrac{1}{y} ;
\end{equation*}

\begin{equation*}
  f(y) = \cos(y) : - \dfrac{d \cos(y)}{d y} \cdot \dfrac{1}{\cos(y)} = \tan(y) .
\end{equation*}
}, such that $ \lambda_{i}(\dot{\gamma_{i}}) = - \frac{d (1 - f(\dot{\gamma_{i}}))}{d \dot{\gamma_{i}}} \cdot \frac{1}{1 - f(\dot{\gamma_{i}})}$ is unbounded above at the zeros, $f(\dot{\gamma_{i}}) \rightarrow 1$. Hence, it implies the following differential form:

\begin{equation*}
  \begin{array}{rcl}
    \displaystyle\lim_{\dot{\gamma_{i}} \rightarrow 1} | \lambda_{i}(\dot{\gamma_{i}} + \Delta \dot{\gamma_{i}}) - \lambda_{i}(\dot{\gamma_{i}})| = \infty & \rightarrow & \lambda_{i}(\dot{\gamma_{i}}) = - \dfrac{d (1 - f(\dot{\gamma_{i}}))}{d \dot{\gamma_{i}}} \cdot \dfrac{1}{1 - f(\dot{\gamma_{i}})} ; \\
    \\
    \lambda_{i}(\dot{\gamma_{i}}) \mathrm{d}\dot{\gamma_{i}} &=& \dfrac{d f(\dot{\gamma_{i}})}{1 - f(\dot{\gamma_{i}})} . \\
  \end{array}
\end{equation*}

Integrating on the full domain $\dot{\gamma_{i}} \in (0,1]$, we obtain a parametric representation of the audit intensity $\Lambda_{i}(0,\dot{\gamma_{i}})$ in terms of $f(\dot{\gamma_{i}})$ equal to

\begin{equation*}
  \begin{array}{rcl}
    \dot{\gamma_{i}} \in (0,1] : \Lambda_{i}(0,\dot{\gamma_{i}}) = \displaystyle\int_{\rightarrow 0^{+}}^{1} \, \lambda_{i}(y) \, \mathrm{d}y &=& \displaystyle\int_{\rightarrow 0^{+}}^{1} \, \dfrac{d f(y)/d y}{1 - f(y)} \, \mathrm{d}y \\
    \\
    &=& - \ln(1 - f(\dot{\gamma_{i}})) \\
  \end{array}
\end{equation*}

\noindent which satisfies the unboundedness condition. Therefore, the total probability of tax audit becomes $\mathbb{P}(q(\dot{\gamma_{i}}) > 0, \lambda_{i}(\dot{\gamma_{i}})) = 1 - e^{-\Lambda_{i}(0,\dot{\gamma_{i}})} = 1 - e^{\ln(1 - f(\dot{\gamma_{i}}))} = f(\dot{\gamma_{i}})$. 

\subsection*{Derivation of Eq. \ref{dp d dot gamma}} \label{derivation dp d dt gamma}

To simplify the analysis, we adopt in this section the prime notation for the first and second derivatives as follows: for the differentiable function $f(\dot{\gamma_{i}}) = f_{\dot{\gamma_{i}}}$, the first and second derivatives regarding the variable $\dot{\gamma_{i}}$ are respectively equal to $d f_{\dot{\gamma_{i}}} / d \dot{\gamma_{i}} = f_{\dot{\gamma_{i}}}^{\prime}$, $d^{2} f_{\dot{\gamma_{i}}} / d \dot{\gamma_{i}}^{2} = f_{\dot{\gamma_{i}}}^{\prime \prime}$.

Differentiating the optimal transfer price $p^{*} (\dot{\gamma_{i}}, \cdot)$ with respect to the variable $\dot{\gamma_{i}}$ as parametrisation in Section \ref{Modelling a General Tax Enforcement Effect}, we obtain the following standard form:

\begin{equation*}
  \begin{array}{rcl}
    \dfrac{\partial p^{*} (\dot{\gamma_{i}}, \cdot)}{\partial \dot{\gamma_{i}}} &=& \dfrac{\partial \bar{p}^{c}}{\partial \dot{\gamma_{i}}} + \dfrac{\partial \Delta p^{*}}{\partial \dot{\gamma_{i}}} \\
    \\
    &=& 0 + \dfrac{\partial}{\partial \dot{\gamma_{i}}} \left( \left( \dfrac{|\tau_{2} - \tau_{1}|}{f_{\dot{\gamma_{i}}} \cdot (1 + z_{i}) \cdot \tau_{i} \cdot \left( 1 + \frac{1}{g_{\dot{\gamma_{i}}}} \right)} \right)^{g_{\dot{\gamma_{i}}}} (p^{c} - \bar{p}^{c}) \right) \\
    \\
    &=& \Delta p^{*} \cdot \left[ g_{\dot{\gamma_{i}}}^{\prime} \cdot \ln \left( \dfrac{|\tau_{2} - \tau_{1}|}{f_{\dot{\gamma_{i}}} \cdot (1 + z_{i}) \cdot \tau_{i} \cdot \left( 1 + \frac{1}{g_{\dot{\gamma_{i}}}} \right)} \right) + g_{\dot{\gamma_{i}}} \cdot \dfrac{\partial}{\partial \dot{\gamma_{i}}} \ln \left( \dfrac{|\tau_{2} - \tau_{1}|}{f_{\dot{\gamma_{i}}} \cdot (1 + z_{i}) \cdot \tau_{i} \cdot \left( 1 + \frac{1}{g_{\dot{\gamma_{i}}}} \right)} \right) \right] \\
    \\
    &=& \Delta p^{*} \cdot \left[ g_{\dot{\gamma_{i}}}^{\prime} \cdot \ln \left( \dfrac{|\tau_{2} - \tau_{1}|}{f_{\dot{\gamma_{i}}} \cdot (1 + z_{i}) \cdot \tau_{i} \cdot \left( 1 + \frac{1}{g_{\dot{\gamma_{i}}}} \right)} \right) + f_{\dot{\gamma_{i}}}(1 + g_{\dot{\gamma_{i}}}) \cdot \dfrac{g_{\dot{\gamma_{i}}}^{\prime} f_{\dot{\gamma_{i}}} - g_{\dot{\gamma_{i}}} f_{\dot{\gamma_{i}}}^{\prime} - g_{\dot{\gamma_{i}}}^{2} f_{\dot{\gamma_{i}}}^{\prime}}{(f_{\dot{\gamma_{i}}}(1 + g_{\dot{\gamma_{i}}}))^{2}} \right] \\
    \\
    &=& \Delta p^{*} \cdot \left[ g_{\dot{\gamma_{i}}}^{\prime} \cdot \left( \ln \left( \dfrac{|\tau_{2} - \tau_{1}|}{f_{\dot{\gamma_{i}}} \cdot (1 + z_{i}) \cdot \tau_{i} \cdot \left( 1 + \frac{1}{g_{\dot{\gamma_{i}}}} \right)} \right) + \dfrac{1}{1 + g_{\dot{\gamma_{i}}}} \right) - f_{\dot{\gamma_{i}}}^{\prime} \cdot \dfrac{g_{\dot{\gamma_{i}}}}{f_{\dot{\gamma_{i}}}} \right] . \\
  \end{array}   
\end{equation*}

\bibliographystyle{apacite}
\bibliography{simplemodel}

\begin{thebibliography}{}

\bibitem [\protect \citeauthoryear {%
Allingham%
\ \BBA {} Sandmo%
}{%
Allingham%
\ \BBA {} Sandmo%
}{%
{\protect \APACyear {1972}}%
}]{%
sandmo1972}
\APACinsertmetastar {%
sandmo1972}%
\begin{APACrefauthors}%
Allingham, M\BPBI G.%
\BCBT {}\ \BBA {} Sandmo, A.%
\end{APACrefauthors}%
\unskip\
\newblock
\APACrefYearMonthDay{1972}{}{}.
\newblock
{\BBOQ}\APACrefatitle {Income tax evasion: A theoretical analysis} {Income tax
  evasion: A theoretical analysis}.{\BBCQ}
\newblock
\APACjournalVolNumPages{}{1}{3-4}{323--338}.
\PrintBackRefs{\CurrentBib}

\bibitem [\protect \citeauthoryear {%
Alm%
}{%
Alm%
}{%
{\protect \APACyear {2012}}%
}]{%
alm2012}
\APACinsertmetastar {%
alm2012}%
\begin{APACrefauthors}%
Alm, J.%
\end{APACrefauthors}%
\unskip\
\newblock
\APACrefYearMonthDay{2012}{}{}.
\newblock
{\BBOQ}\APACrefatitle {Measuring, explaining, and controlling tax evasion:
  lessons from theory, experiments, and field studies} {Measuring, explaining,
  and controlling tax evasion: lessons from theory, experiments, and field
  studies}.{\BBCQ}
\newblock
\APACjournalVolNumPages{International Tax and Public Finance}{19}{1}{54--77}.
\PrintBackRefs{\CurrentBib}

\bibitem [\protect \citeauthoryear {%
Bartelsman%
\ \BBA {} Beetsma%
}{%
Bartelsman%
\ \BBA {} Beetsma%
}{%
{\protect \APACyear {2003}}%
}]{%
bartelsman2003}
\APACinsertmetastar {%
bartelsman2003}%
\begin{APACrefauthors}%
Bartelsman, E\BPBI J.%
\BCBT {}\ \BBA {} Beetsma, R\BPBI M.%
\end{APACrefauthors}%
\unskip\
\newblock
\APACrefYearMonthDay{2003}{}{}.
\newblock
{\BBOQ}\APACrefatitle {Why pay more? Corporate tax avoidance through transfer
  pricing in OECD countries} {Why pay more? corporate tax avoidance through
  transfer pricing in oecd countries}.{\BBCQ}
\newblock
\APACjournalVolNumPages{Journal of Public Economics}{87}{9-10}{2225--2252}.
\PrintBackRefs{\CurrentBib}

\bibitem [\protect \citeauthoryear {%
Becker%
, Davies%
\BCBL {}\ \BBA {} Jakobs%
}{%
Becker%
\ \protect \BOthers {.}}{%
{\protect \APACyear {2017}}%
}]{%
becker2017}
\APACinsertmetastar {%
becker2017}%
\begin{APACrefauthors}%
Becker, J.%
, Davies, R\BPBI B.%
\BCBL {}\ \BBA {} Jakobs, G.%
\end{APACrefauthors}%
\unskip\
\newblock
\APACrefYearMonthDay{2017}{}{}.
\newblock
{\BBOQ}\APACrefatitle {The economics of advance pricing agreements} {The
  economics of advance pricing agreements}.{\BBCQ}
\newblock
\APACjournalVolNumPages{Journal of Economic Behavior \&
  Organization}{134}{}{255--268}.
\PrintBackRefs{\CurrentBib}

\bibitem [\protect \citeauthoryear {%
Bernard%
, Jensen%
\BCBL {}\ \BBA {} Schott%
}{%
Bernard%
\ \protect \BOthers {.}}{%
{\protect \APACyear {2006}}%
}]{%
bernard2006}
\APACinsertmetastar {%
bernard2006}%
\begin{APACrefauthors}%
Bernard, A\BPBI B.%
, Jensen, J\BPBI B.%
\BCBL {}\ \BBA {} Schott, P\BPBI K.%
\end{APACrefauthors}%
\unskip\
\newblock
\APACrefYearMonthDay{2006}{}{}.
\newblock
\APACrefbtitle {Transfer pricing by US-based multinational firms} {Transfer
  pricing by us-based multinational firms}\ \APACbVolEdTR{}{\BTR{}}.
\newblock
\APACaddressInstitution{}{National Bureau of Economic Research (No. w12493)}.
\PrintBackRefs{\CurrentBib}

\bibitem [\protect \citeauthoryear {%
Clausing%
}{%
Clausing%
}{%
{\protect \APACyear {2003}}%
}]{%
clausing2003}
\APACinsertmetastar {%
clausing2003}%
\begin{APACrefauthors}%
Clausing, K\BPBI A.%
\end{APACrefauthors}%
\unskip\
\newblock
\APACrefYearMonthDay{2003}{}{}.
\newblock
{\BBOQ}\APACrefatitle {Tax-motivated transfer pricing and US intrafirm trade
  prices} {Tax-motivated transfer pricing and us intrafirm trade
  prices}.{\BBCQ}
\newblock
\APACjournalVolNumPages{Journal of Public Economics}{87}{9-10}{2207--2223}.
\PrintBackRefs{\CurrentBib}

\bibitem [\protect \citeauthoryear {%
Cristea%
\ \BBA {} Nguyen%
}{%
Cristea%
\ \BBA {} Nguyen%
}{%
{\protect \APACyear {2016}}%
}]{%
cristea2016}
\APACinsertmetastar {%
cristea2016}%
\begin{APACrefauthors}%
Cristea, A\BPBI D.%
\BCBT {}\ \BBA {} Nguyen, D\BPBI X.%
\end{APACrefauthors}%
\unskip\
\newblock
\APACrefYearMonthDay{2016}{}{}.
\newblock
{\BBOQ}\APACrefatitle {Transfer pricing by multinational firms: New evidence
  from foreign firm ownerships} {Transfer pricing by multinational firms: New
  evidence from foreign firm ownerships}.{\BBCQ}
\newblock
\APACjournalVolNumPages{American Economic Journal: Economic
  Policy}{8}{3}{170--202}.
\PrintBackRefs{\CurrentBib}

\bibitem [\protect \citeauthoryear {%
Davies%
, Martin%
, Parenti%
\BCBL {}\ \BBA {} Toubal%
}{%
Davies%
\ \protect \BOthers {.}}{%
{\protect \APACyear {2018}}%
}]{%
davies2018}
\APACinsertmetastar {%
davies2018}%
\begin{APACrefauthors}%
Davies, R\BPBI B.%
, Martin, J.%
, Parenti, M.%
\BCBL {}\ \BBA {} Toubal, F.%
\end{APACrefauthors}%
\unskip\
\newblock
\APACrefYearMonthDay{2018}{}{}.
\newblock
{\BBOQ}\APACrefatitle {Knocking on tax haven’s door: Multinational firms and
  transfer pricing} {Knocking on tax haven’s door: Multinational firms and
  transfer pricing}.{\BBCQ}
\newblock
\APACjournalVolNumPages{Review of Economics and Statistics}{100}{1}{120--134}.
\PrintBackRefs{\CurrentBib}

\bibitem [\protect \citeauthoryear {%
Eden%
}{%
Eden%
}{%
{\protect \APACyear {2001}}%
}]{%
eden2001}
\APACinsertmetastar {%
eden2001}%
\begin{APACrefauthors}%
Eden, L.%
\end{APACrefauthors}%
\unskip\
\newblock
\APACrefYearMonthDay{2001}{}{}.
\newblock
{\BBOQ}\APACrefatitle {Taxes, transfer pricing, and the multinational
  enterprise} {Taxes, transfer pricing, and the multinational
  enterprise}.{\BBCQ}
\newblock
\APACjournalVolNumPages{The Oxford Handbook in International Business, Oxford
  University Press: Oxford}{}{}{591--619}.
\PrintBackRefs{\CurrentBib}

\bibitem [\protect \citeauthoryear {%
Hines~Jr%
\ \BBA {} Rice%
}{%
Hines~Jr%
\ \BBA {} Rice%
}{%
{\protect \APACyear {1994}}%
}]{%
hines1994}
\APACinsertmetastar {%
hines1994}%
\begin{APACrefauthors}%
Hines~Jr, J\BPBI R.%
\BCBT {}\ \BBA {} Rice, E\BPBI M.%
\end{APACrefauthors}%
\unskip\
\newblock
\APACrefYearMonthDay{1994}{}{}.
\newblock
{\BBOQ}\APACrefatitle {Fiscal paradise: Foreign tax havens and American
  business} {Fiscal paradise: Foreign tax havens and american business}.{\BBCQ}
\newblock
\APACjournalVolNumPages{The Quarterly Journal of Economics}{109}{1}{149--182}.
\PrintBackRefs{\CurrentBib}

\bibitem [\protect \citeauthoryear {%
Kant%
}{%
Kant%
}{%
{\protect \APACyear {1988}}%
}]{%
kant1988}
\APACinsertmetastar {%
kant1988}%
\begin{APACrefauthors}%
Kant, C.%
\end{APACrefauthors}%
\unskip\
\newblock
\APACrefYearMonthDay{1988}{}{}.
\newblock
{\BBOQ}\APACrefatitle {Endogenous transfer pricing and the effects of uncertain
  regulation} {Endogenous transfer pricing and the effects of uncertain
  regulation}.{\BBCQ}
\newblock
\APACjournalVolNumPages{Journal of International Economics}{24}{1-2}{147--157}.
\PrintBackRefs{\CurrentBib}

\bibitem [\protect \citeauthoryear {%
Klir%
\ \BBA {} Yuan%
}{%
Klir%
\ \BBA {} Yuan%
}{%
{\protect \APACyear {1995}}%
}]{%
klir1995}
\APACinsertmetastar {%
klir1995}%
\begin{APACrefauthors}%
Klir, G.%
\BCBT {}\ \BBA {} Yuan, B.%
\end{APACrefauthors}%
\unskip\
\newblock
\APACrefYear{1995}.
\newblock
\APACrefbtitle {Fuzzy sets and fuzzy logic} {Fuzzy sets and fuzzy logic}\
  (\BVOL~4).
\newblock
\APACaddressPublisher{}{Prentice hall New Jersey}.
\PrintBackRefs{\CurrentBib}

\bibitem [\protect \citeauthoryear {%
Levaggi%
\ \BBA {} Menoncin%
}{%
Levaggi%
\ \BBA {} Menoncin%
}{%
{\protect \APACyear {2013}}%
}]{%
levaggi2013}
\APACinsertmetastar {%
levaggi2013}%
\begin{APACrefauthors}%
Levaggi, R.%
\BCBT {}\ \BBA {} Menoncin, F.%
\end{APACrefauthors}%
\unskip\
\newblock
\APACrefYearMonthDay{2013}{}{}.
\newblock
{\BBOQ}\APACrefatitle {Optimal dynamic tax evasion} {Optimal dynamic tax
  evasion}.{\BBCQ}
\newblock
\APACjournalVolNumPages{Journal of Economic Dynamics and
  Control}{37}{11}{2157--2167}.
\PrintBackRefs{\CurrentBib}

\bibitem [\protect \citeauthoryear {%
Nielsen%
, Schindler%
\BCBL {}\ \BBA {} Schjelderup%
}{%
Nielsen%
\ \protect \BOthers {.}}{%
{\protect \APACyear {2014}}%
}]{%
nielsen2014}
\APACinsertmetastar {%
nielsen2014}%
\begin{APACrefauthors}%
Nielsen, S.%
, Schindler, D.%
\BCBL {}\ \BBA {} Schjelderup, G.%
\end{APACrefauthors}%
\unskip\
\newblock
\APACrefYearMonthDay{2014}{}{}.
\newblock
\APACrefbtitle {Abusive transfer pricing and economic activity} {Abusive
  transfer pricing and economic activity}\ \APACbVolEdTR{}{\BTR{}}.
\newblock
\APACaddressInstitution{}{CESifo Working Paper No. 4975}.
\PrintBackRefs{\CurrentBib}

\bibitem [\protect \citeauthoryear {%
OECD%
}{%
OECD%
}{%
{\protect \APACyear {2017}}%
}]{%
oecd2017}
\APACinsertmetastar {%
oecd2017}%
\begin{APACrefauthors}%
OECD.%
\end{APACrefauthors}%
\unskip\
\newblock
\APACrefYear{2017}.
\newblock
\APACrefbtitle {OECD Transfer Pricing Guidelines for Multinational Enterprises
  and Tax Administrations 2017} {Oecd transfer pricing guidelines for
  multinational enterprises and tax administrations 2017}.
\newblock
\begin{APACrefURL}
  \url{https://www.oecd-ilibrary.org/content/publication/tpg-2017-en}
  \end{APACrefURL}
\newblock
\begin{APACrefDOI} \doi{https://doi.org/https://doi.org/10.1787/tpg-2017-en}
  \end{APACrefDOI}
\PrintBackRefs{\CurrentBib}

\bibitem [\protect \citeauthoryear {%
Overesch%
}{%
Overesch%
}{%
{\protect \APACyear {2006}}%
}]{%
overesch2006}
\APACinsertmetastar {%
overesch2006}%
\begin{APACrefauthors}%
Overesch, M.%
\end{APACrefauthors}%
\unskip\
\newblock
\APACrefYearMonthDay{2006}{}{}.
\newblock
\APACrefbtitle {Transfer pricing of intrafirm sales as a profit shifting
  channel-Evidence from German firm data} {Transfer pricing of intrafirm sales
  as a profit shifting channel-evidence from german firm data}\
  \APACbVolEdTR{}{\BTR{}}.
\newblock
\APACaddressInstitution{}{ZEW Discussion Papers No. 06-84}.
\PrintBackRefs{\CurrentBib}

\bibitem [\protect \citeauthoryear {%
Swenson%
}{%
Swenson%
}{%
{\protect \APACyear {2001}}%
}]{%
swenson2001}
\APACinsertmetastar {%
swenson2001}%
\begin{APACrefauthors}%
Swenson, D\BPBI L.%
\end{APACrefauthors}%
\unskip\
\newblock
\APACrefYearMonthDay{2001}{}{}.
\newblock
{\BBOQ}\APACrefatitle {Tax reforms and evidence of transfer pricing} {Tax
  reforms and evidence of transfer pricing}.{\BBCQ}
\newblock
\APACjournalVolNumPages{National Tax Journal}{}{}{7--25}.
\PrintBackRefs{\CurrentBib}

\bibitem [\protect \citeauthoryear {%
Verdegay%
}{%
Verdegay%
}{%
{\protect \APACyear {1982}}%
}]{%
verdegay1982}
\APACinsertmetastar {%
verdegay1982}%
\begin{APACrefauthors}%
Verdegay, J\BPBI L.%
\end{APACrefauthors}%
\unskip\
\newblock
\APACrefYearMonthDay{1982}{}{}.
\newblock
{\BBOQ}\APACrefatitle {Fuzzy mathematical programming} {Fuzzy mathematical
  programming}.{\BBCQ}
\newblock
\APACjournalVolNumPages{Fuzzy information and decision processes}{231}{}{237}.
\PrintBackRefs{\CurrentBib}

\bibitem [\protect \citeauthoryear {%
Zadeh%
\ \protect \BOthers {.}}{%
Zadeh%
\ \protect \BOthers {.}}{%
{\protect \APACyear {1965}}%
}]{%
zadeh1965}
\APACinsertmetastar {%
zadeh1965}%
\begin{APACrefauthors}%
Zadeh, L\BPBI A.%
\BCBT {}\ \BOthersPeriod {.}
\end{APACrefauthors}%
\unskip\
\newblock
\APACrefYearMonthDay{1965}{}{}.
\newblock
{\BBOQ}\APACrefatitle {Fuzzy sets} {Fuzzy sets}.{\BBCQ}
\newblock
\APACjournalVolNumPages{Information and control}{8}{3}{338--353}.
\PrintBackRefs{\CurrentBib}

\bibitem [\protect \citeauthoryear {%
Zimmermann%
}{%
Zimmermann%
}{%
{\protect \APACyear {1991}}%
}]{%
zimmermann1991}
\APACinsertmetastar {%
zimmermann1991}%
\begin{APACrefauthors}%
Zimmermann, H\BHBI J.%
\end{APACrefauthors}%
\unskip\
\newblock
\APACrefYear{1991}.
\newblock
\APACrefbtitle {Fuzzy Set Theory -- and Its Applications} {Fuzzy set theory --
  and its applications}.
\newblock
\APACaddressPublisher{}{Kluwer Academic Publishers, 2nd, revised edition}.
\PrintBackRefs{\CurrentBib}

\end{thebibliography}

\end{document}